\documentclass[12pt,preprint]{aastex}

\shorttitle{Ly$\alpha$ emission on GRB spectrum}
\shortauthors{Xu, \& Wu}

\begin{document}

\title{On the formation of Lyman $\alpha$ emission from resonantly scattered continuum photons of GRB's afterglow}
\author{Wen Xu\altaffilmark{1}, Xiang-Ping Wu\altaffilmark{1}}

\altaffiltext{1}{National Astronomical Observatories, Chinese Academy of Sciences, Beijing 100012,China}

\begin{abstract}

\vspace{0cm}

The continuum spectrum of GRB's afterglow at Lyman $\alpha$ wavelength is known to be otherwise featureless
except the existence of a pair of smooth damping wings. Resonant scattering of photons
with the ambient neutral hydrogen around GRB may alter this picture. We study the formation and
evolution of the spectral imprint of these resonantly scattered
photons in the context of GRB's afterglow.
  Based on an  analytic model that includes photons which are scattered only once,
 as well as a complete treatment of all the scatterings  using Monte Carlo simulations, we are able to 
calculate the spectrum and luminosity of this Lyman $\alpha$ emission from a very early moment on to a late epoch.
 We find that the amount, the motion
 and the geometry of the neutral hydrogen
around GRB, together with the time behavior of the source are the crucial factors which affect
  the predicted luminosity and spectral profile.    
The flux of the Ly$\alpha$ emission is found to be mainly contributed by photons which are scattered only once. The flux is of order  $10^{-4}$ to $10^{-9}$ relative to the undecayed maximum flux of the transmitted continuum, making the feature negligible but potentially observable. If not obscured by host galaxy's DLA or intergalactic neutral hydrogen, the feature may appear sometime
from one hour to several years when
the directly transmitted light has faded away.
 This scattered emission feature can be distinguished from Ly$\alpha$ photons of other origins by its luminosity evolution, and 
by its gradual narrowing of 
profile with time.  
The typical time scale for spectral variance is that of the light crossing time of a hydrogen clump close to the GRB.
If observed, the resonant peaks' time dependent behavior is a scanning probe on the distribution of neutral hydrogen
in GRB's  immediate neighborhood.

\end{abstract}


\section{Introduction}

Gamma-ray bursts (GRB) are the most energetic explosions of astrophysical objects known to human being
(see reviews of Piran 2004; Meszaros 2006; Woosley \& Bloom 2006). 
GRB cosmology has been part of the hierarchical structure formation paradigm of the $\Lambda$CDM model (Xu \& Fang 1999; Bromm \& Loeb 2007; Nagamine, Zhang \& Hernquist 2008; Pontzen et al. 2009).
With the recent discovery of z=8.3 GRB 090423 (Tanvir et al. 2009, Salvaterra et al. 2009),
GRB's Ly$\alpha$ damping wings are becoming a powerful tool to probe chemical evolution, star formation, the reionization era and the dark ages
(Kawai et al. 2006; Totani et al. 2006; Gallerani et al. 2008; Mesinger \& Furlanetto 2008; McQuinn et al. 2008).

However,  the damping wings are not a clean probe. 
There are at least two kinds of damping wings. The first one is caused by the scattering away of photons by IGM's
neutral hydrogen atoms along line of sight(Miralda-Escude 1998), the optical depth of which follows an integration over the Voigt wings of the scattering cross section function and thus can be approximated as inversely proportional to $\Delta \lambda$. The second kind of damping wing is caused  by scattering from a local DLA cloud (Totani 2006), which maps directly the Voigt profile and is roughly
 proportional to $\frac{1}{(\Delta \lambda) ^2}$. Besides the uncertain profile of the absorptions,
Ly$\alpha$ emissions have been observed to be present in the centers of damped Lyman $\alpha$ absorption spectra of
QSO-pDLAs (\underline{p}roximate \underline{D}amped \underline{L}yman $\alpha$ \underline{A}bsorption systems, e.g. Hunstead, Pettini, \& Fletcher 1990; Leibundgut \& Robertson 1999; Moller et al. 2002; Hennawi et al. 2009), as well as in those of GRB-hDLAs (\underline{h}ost DLA) (e.g. Vreeswijk et al. 2004; Totani et al. 2006). These Ly$\alpha$ photons may
have a number of origins. They can be produced
 by recombination in star formation regions,  by recombination in AGN powered ionization,  by fluorescence of gas cloud illuminated by a nearby QSO, by gravitational heating in cooling streams, or by resonant scattering of the continuum afterglow of GRBs as investigated in this paper. 

The escape of Lyman $\alpha$ photons through an optical thick cloud has been studied by many authors 
(Osterbrock 1962; Adams 1972; Urbaniak \& Wolfe 1981; Loeb \& Rybicki 1999; Zheng \& Miralda-Escude 2002; Tasitsiomi 2006; Verhamme et al. 2006;
 Laursen \& Sommer-Larsen 2007; Dijkstra \& Loeb 2008; Pierleoni et al, 2009; Roy et al. 2009a). In the context of GRB's afterglow,
resonantly scattered photons of GRB's continuum optical light are not lost.
 They are retained in neutral hydrogen clouds and will arrive observers at a later time.
 These scattered and thus delayed photons may look brighter than the transmitted ones because GRB's optical light decays fast.
 Thus resonant scattering is a new mechanism  which produces weak Ly$\alpha$ emission features in GRB's spectrum.

In \S2, the modeling and the physics of resonant scattering of Ly$\alpha$ photons are reviewed.
In \S 3, we use a simplified but analytic model to illustrate how emission features can be formed when continuum photons
at Lyman $\alpha$ wavelength collide resonantly with circumburst neutral hydrogen clouds.
In \S 4, we model the complete scattering process with Monte Carlo (MC) simulations.  
The effects of model parameters and the observability are discussed and concluded in \S 5.

\section{Resonant scattering at Lyman $\alpha$ frequency} 

Resonant scattering of Ly$\alpha$ photons in a cosmic setting
 has been studied by many authors employing either Monte Carlo simulations 
(e.g. Loeb \& Rybicki 1999; Zheng \& Miralda-Escude 2002; Tasitsiomi 2006; Verhamme et al. 2006; Laursen \& Sommer-Larsen 2007; Dijkstra \& Loeb 2008; Pierleoni et al, 2009),
or  the radiative transfer equation(Roy et al. 2009abc).  We refer to Roy et al. (2009a) for notations and conventions used in this paper. 

The resonant scattering cross section is (e.g. Gunn \& Peterson 1965)
$
\sigma(\nu)=\sigma_0 g(\nu-\nu_{\alpha})
$
in which $\sigma_0=\frac{\pi e^2 f}{m_e c} $,
 $f=0.416$, $\nu_{\alpha}=2.46\times10^{15} s^{-1}$, $g(\nu-\nu_{\alpha})$ is the normalized line profile
$
1=\int^{\infty}_{-\infty} g(\nu-\nu_{\alpha})d\nu$.
If we introduce a dimensionless frequency $x\equiv\frac{\nu-\nu_{\alpha}}{\Delta \nu_D}$,
where $\Delta \nu_D= \frac{V_D}{c}\nu_0 = 1.06\times 10^{11}(\frac{V_D}{12.9 \rm{kms}^{-1}})$Hz and
$V_D$ is the Doppler velocity.  The value $12.9 \rm{kms}^{-1}$ corresponds to a temperature  of $10^4 K$ in a static medium. However, in absorbing gas 
temperature is not the major source of Doppler motion. $V_D$ is more likely to be contributed by macroscopic motions rather than thermal motions.
Our results are not sensitive to $V_D$ because the interested scatterings
  happen at wing frequencies of the resonant line($x \sim 50$). 
With these notations,
\begin{equation}
\sigma(x)=\sigma_0 \phi(x) (\Delta \nu_D)^{-1}
\end{equation}
where the normalized Voigt profile is(see, e.g., Hummer 1962, eq2.22.1)
\begin{equation}
\phi(x)=\frac{a}{\pi^{\frac{3}{2}}}\int^{\infty}_{-\infty}\frac{e^{-y^2}}{(x-y)^2+a^2}dy
\end{equation}
which is the joint effect of the Gaussian distribution of thermal velocity of neutral hydrogen atom
 and the Lorentz profile of cross section in the rest frame of the atom. It is 
normalized as
$
1=\int^{\infty}_{-\infty} \phi(x)dx
$.
$a=\frac{\Lambda}{4\pi \Delta \nu_D}=4.70\times 10^{-4} (\frac{12.9 \rm{kms}^{-1}}{V_D})$ is a shape parameter in
 line profile (Hummer 1962), where $\Lambda=6.25 \times 10^8 s^{-1}$ (see, e.g., Miralda-Escude \& Rees 1998)
 is the total decay constant for the Ly$\alpha$ resonance. 
 
A Lyman $\alpha$ photon at frequency $x$ will experience a free path length $l$
 before it scatters resonantly with a neutral hydrogen (HI) atom.
 The distribution of length $l$ follows $e^{-\frac{l}{l_*}}$ where $l_*$  is the mean length of free path  
$l_*= \frac{1}{n\sigma(x)}$. The optical depth incurred over a segment of light path $dl$ is
$
d\tau=n\sigma dl
$.
The total optical depth for a cloud of column density $N_{HI}$ is
$
\tau(x)  = N_{HI} \sigma(x) = \tau_0 \phi(x)
$
,where 
$
\tau_0 = N_{HI} \sigma_0 (\Delta \nu_D)^{-1} =
 1.04\times10^{7}(\frac{12.9 \rm{kms}^{-1}}{V_D})\left (\frac{N_{HI}}{10^{20}cm^{2}}\right)
$
. Therefore, the optical depth at the Lyman $\alpha$ line center frequency is
$
\tau(0)=\tau_0\cdot \frac{1}{\sqrt{\pi}}=5.86\times10^6 (\frac{12.9 \rm{kms}^{-1}}{V_D})\left (\frac{N_{HI}}{10^{20}cm^{2}}\right)
$.

To study the details of resonant scattering of a Lyman $\alpha$ photon with a HI atom,
 we follow Field (1959)'s scattering geometry and notations.
 The coordinates are chosen in such a way that the incoming photon is in z direction and
the unit vector of HI atom's velocity before scattering is
$\hat{e}_{\mathbf{V}}=\mathrm{sin}\eta \hspace{1mm}\hat{e}_x + \mathrm{cos}\eta \hspace{1mm} \hat{e}_z$
where $\eta$ is the angle between the incoming photon and the direction of the motion of HI atom.
The unit vector of scattered photon can be expressed as
$  \hat{e}_{photon'}=\mathrm{sin}\theta\mathrm{cos}\phi\hspace{1mm} \hat{e}_x 
+\mathrm{sin}\theta\mathrm{sin}\phi\hspace{1mm} \hat{e}_y +
\mathrm{cos}\theta\hspace{1mm} \hat{e}_z$,
 where $\theta$ and $\phi$ are the angles of the outgoing photon in spherical coordinates. 
The dimensionless projected velocity of atom along the direction of incoming photon is
$ v_{//} = \frac{V}{V_D} \hspace{2mm} \hat{e}_{\mathbf{V}} \cdot \hat{e}_{photon}$ where $V_D$ is the Doppler velocity.

The incoming photon of frequency x has an effective frequency $ \tilde{x}=x-v_{//} $
 when translated into the rest reference system of the hydrogen atom.
Using the notation of $x'$ to represent the laboratory frequency of the outgoing photon,
\begin{eqnarray*}
x' &=& x- v \hat{e}_{\mathbf{V}} \cdot \hat{e}_{photon} +  v \hat{e}_{\mathbf{V}} \cdot \hat{e}_{photon'} + \frac{h\nu_0}{Mc V_D}( \hat{e}_{z} -\hat{e}_{photon'}) \cdot \hat{e}_{photon'}\\
   &=& x-v{\rm cos}\eta + v {\rm cos}\eta {\rm cos}\theta + v {\rm sin} \theta {\rm sin} \eta {\rm cos}\phi - b(1-\rm{cos}\theta)
\end{eqnarray*}
\begin{equation}
\end{equation}
where $b=\frac{h\nu_0}{m_e V_{D} c}$ is the recoil parameter. We adopt $b=0.03$ in this paper.
By recoil the HI atom gets a velocity increment of
$\frac{h\nu_0}{Mc}( \hat{e}_{z} -\hat{e}_{photon'})$, causing an additional term of
 $\frac{h\nu_0}{Mc V_D}( \hat{e}_{z} -\hat{e}_{photon'}) \cdot \hat{e}_{photon'}$ in the $x'$ formula. 
This term has two contributions.
 The reference system gets a backward velocity increment 
against the direction of re-emitted photon, which is $-\frac{h\nu_0}{Mc} \hat{e}_{photon'}$. 
Besides,
HI atom has gained a velocity increment of $\frac{h\nu}{Mc} \hat{e}_{z}$ when absorbing the incoming photon.

\begin{figure} %[TB]
\begin{center}
\includegraphics[height=6.5 cm]{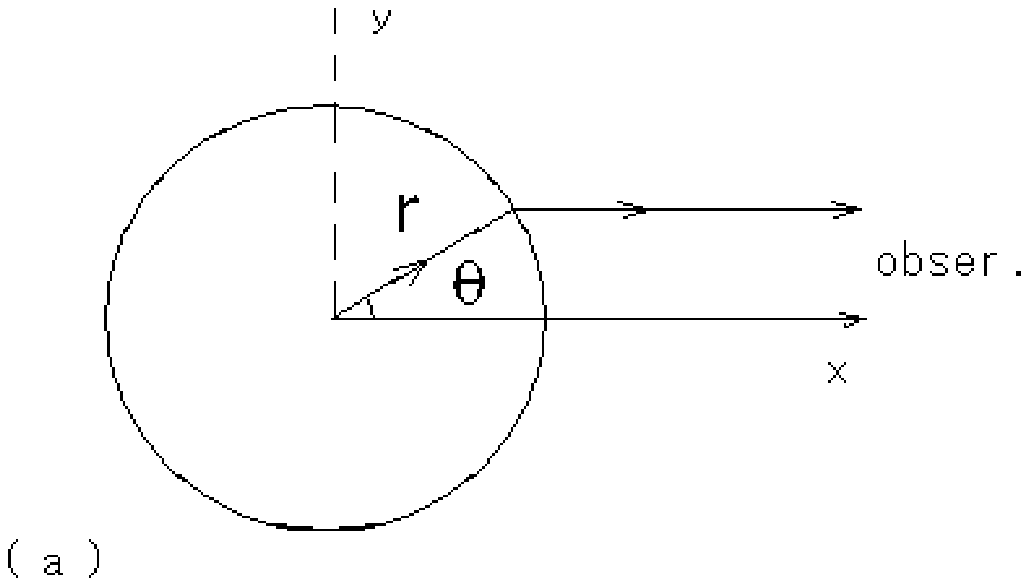}
\includegraphics[height=6.5 cm]{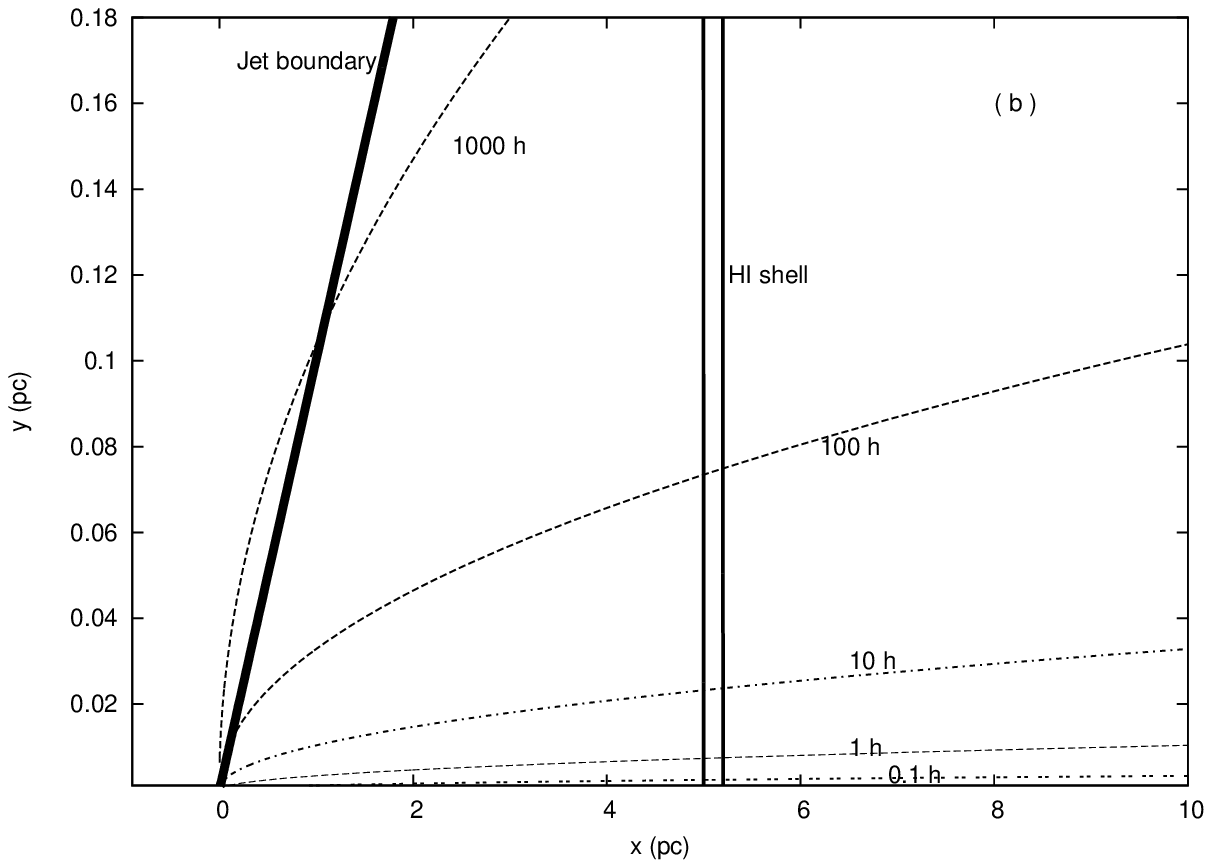}
\caption{Geometry of the resonant scattering around a GRB in the Shell model. (a) light path of a scattered photon. (b) distribution of time delays for photons experiencing a single collision. x axis is along the line of sight. y axis is a direction perpendicular to the line of sight. The numbers next to the curves  correspond to time delays of these curves where the scatterings happen and are measured at the GRB reference system. The location of the thin shell is shown at 5pc from GRB. Although an isotropic emission is treated, a jet boundary of opening angle 0.1 radian is shown as an example. }
\label{fig1}
\end{center}
\end{figure}

In adopting model parameters for GRB's afterglow, we consider a low redshift GRB with optical flash, but without any host galaxy DLA
 so that the scattering effects at GRB's immediate neighborhood can be isolated from other effects along line-of-sight. 
We employ spherically symmetric representations of small cloudlets around GRB.
Two models of the neutral hydrogen distribution are considered. 
 In the Sphere model,   a uniform host cloud surrounds the GRB. (We also extend our equations to be  applicable to
  a polytrope whose density is a function of radius.)
In the Shell model,
 a shell-like distribution of gas intercepts photons. Hydrogen  is uniformly distributed
 on the shell in the Shell model.
In Fig. 1 we illustrate the scattering geometry of the Shell model.  A thin shell of  $N_{HI}$ 
is at 5pc away from GRB with a shell thickness of 0.065pc and HI number density $n \sim 10^3 \rm{cm}^3$.
 This same figure also illustrates our Sphere model, taking the shell position as the outer surface of the cloud. Shell or cloud sizes from 0.01pc to 1 kpc have been studied.  
 Although GRB afterglows are consistent with a scenario in which 
every afterglow is beamed, we tried to avoid the added complexity of beam angle by treating the radiation isotropically. Because of the finite
speed of light following {\it Special Relativity}, the afterglow photons actually don't feel the existence of a beam boundary at small times before 
jet break happens.
 We further assume that the neutral gas is at rest in hydrodynamical equilibrium following Maxwellian velocity distribution.
  A column density of $N_{HI}=10^{20.3} \rm{cm}^{-2}$ is used, which is at the lower end of the observed GRB-DLAs .
 Fig. 1b shows the time delay distribution for the scattered-once photons. This time delay is purely geometric and not affected by how neutral hydrogen  is 
distributed, or by its total amount, thus applies to both models.   

At the site of GRB, we model the photon source function $s(t)$ as an initial plateau with
 a single power law function cut off.
$s(t) =  \left(\frac{t-t_{trig}}{t_s}\right)^{-\alpha}$ if $ t > t_s +t_{trig}$, and $s(t)=1$  if $ t_{trig}< t < t_s +t_{trig}$. 
$\alpha$ is observed to be between 0.5 and 3 in real GRBs.
 %Among the Swift afterglow, the initial X-ray decay is  $\alpha =3- 5$.
 We adopt $\alpha=2$ corresponding 
to $\Gamma=t^{-1/2}$ (Rhoads 1999). $t_s=50$ sec is the burst duration at the source.
A flat featureless spectrum of GRB is assumed 
across the line profile. For simplicity, we assume that the initial continuum flux is unitary (=1) for the unit system we use.

\section{Analytic Modeling of resonant scattering around GRB}

In a static medium, the radiative transfer equation for a pencil of photons is (Chandrasekhar 1950, Chapter 1, Eq.(49))
\begin{equation}
\frac{d J(x,\tau)}{d \tau}=-\phi(x) J(x,\tau) + \Im(x,\tau)
\end{equation}
where $J(x,\tau)=I_{\nu}/h\nu$ is the specific number density of photons, $I_{\nu}$
 is the specific intensity (Chandrasekhar 1950, chapter , Eq.(1)) and $\Im$ is  the source function.

In the traditional way of discussing damping wings the source term is ignored, 
\begin{equation}
\frac{d J_0(\tau,x)}{d \tau} = -\phi(x) J_0(\tau,x).
\end{equation}
The observed flux directly from GRB is $ f_0(x,\frac{t_{obs}}{1+z}) h\nu_{obs} \hspace{3mm} \rm{erg cm}^{-2}s^{-1}$. Thus the number flux 
in the GRB's redshift frame is
\begin{equation}
f_0(x,t) =\Sigma J_0(x,\tau)= f_{max} \cdot s(t) \cdot e^{-\tau_{0} \phi(x)} \hspace{2mm} \rm{cm}^{-2}s^{-1}
\end{equation}
in which $\Sigma$ represents summation over all the pencils of photons arriving in unit area at the observer. Thus, 
$f_{max}=\frac{F(x)}{4\pi d_L^2}$ is the flux of the source if it were not decaying (thus it is the maximum).
$F(x)$ is the photon release rate at the source in unit of photons per $x$ per second, 
We conveniently choose  $f_{max}=1 $ to illustrate. It is
 in unit of photons per x per second. $s(t)$ is the source function introduced in \S 2. $t$ and $x$ refer to values at the source.
The number flux $f$ is related to the conventional definition by
$f_{\lambda}(\lambda_{obs},t_{obs}) = \frac{\nu_0}{\Delta\nu_D \lambda_0} f(x,\frac{t_{obs}}{1+z}) h\nu_{obs} \hspace{2mm}$ erg cm $^{-2}s^{-1}\AA^{-1}$.

Eq(5) is inaccurate because the scattered photons may be
scattered back into the line of sight. The scattered back photons can be described by the source term $\Im$. 
The accurate flux can always be written as $J\equiv J_0+J_s$.
From Eq(4) we have
\begin{equation}
\frac{d J_s(x,\tau)}{d \tau}=-\phi(x) J_s(x,\tau) + \Im(\tau,x)
\end{equation}
We use $\Im_1$ to name the contribution from the scattering of photons  directly from the photon source. Similarly,
$\Im_n$ denotes the contribution to the photon flux from 
the resonant scattering of photons which have been scattered $n-1$ times ($n\geq 2$). 
In a medium where there is no explicit photon source, the only contribution to the source function $\Im$ is from resonant scattering.
Therefore, $\Im=\Sigma_{i=1}^{+\infty} \Im_i$.
On the other hand, the intensity of photon flux of scattered component
 can always be formally expanded as $J_s = \Sigma_{n=1}^{+\infty} J_n$ if we define $J_n$ as the flux of photons
 which are scattered exactly n times, 
\begin{equation}
\frac{d J_n(x,\tau)}{d \tau}=-\phi(x) J_n(x,\tau) +\Im_n
\end{equation}
Specifically,
\begin{equation}
\frac{d J_1(x,\tau)}{d \tau}=-\phi(x) J_1(x,\tau) + \Im_1
\end{equation}

So the traditional damping wing is  the zeroth order approximation along  a  
 perturbative approach in which  photons of any times of scattering will be included.
In this section, we go one step further to include photons
which have scattered for only once, and ignore photons contributed from multiple scatterings.
The advantage is the ability to include the scattering geometry analytically and to show how the
basic scatter feature is generated and scaled. In next section, we will show by Monte Carlo simulation method that the scattered emission is indeed dominated by photons which are scattered once, when the observation time is small.

For photons which are scattered once, we can ignore the tiny transfer in frequency which is of order $x \sim 1$.
Thus under assumption of elastic scatterings, the scattered photons have the same
frequency as what they come with.
For the Shell model under thin shell approximation ($dr <<r $) from Eq(9), the number flux of the scattered-once light is
\begin{equation}
f_1(x,t) =\Sigma J_1(x,\tau)= \frac{f_{max}}{2} \cdot \int^{\theta_*}_0 \theta d\theta \cdot s(t-\frac{r}{c}(1-cos\theta)) \cdot \tau_{1}\phi(x) e^{-\tau_{1} \phi(x)} 
\end{equation}
in unit of  $\rm{ cm}^{-2}s^{-1}$ where $\Sigma$ represents summation over all the pencils of scattered once photons which
arrive at the observer within unit area . This equation is accurate under these assumptions but
have larger errors if the jet boundary $\theta_* >\frac{\pi}{2}$ when the photons scattered from the farther half of the shell have to
cross the front shell to reach the observer.$\tau_1$ is the optical depth of the shell along the actual light path (Fig. 1a). For spherically symmetric medium, $\tau_1=\tau_0$. Since the  directly arriving photons and the scattered photons follow different light paths, it's possible 
that $\tau_1$ may be different from $\tau_0$.

 For the Sphere model of HI distribution,  the scattered-once component is 
\begin{eqnarray*}
f_1(x,t) & =&   \frac{(n+1)f_{max}}{2} 
\int^{R}_{0} \left(\frac{r}{R}\right)^{-n}\frac{dr}{R}\int^{\theta_*}_0 sin\theta d\theta 
\cdot s\left( t - \frac{r}{c} (1-cos\theta) \right ) \\
& & \cdot \tau_{1}\phi(x) e^{-\frac{\tau_{1}\phi(x)}{R}\left(r+
\sqrt{R^2-r^2sin^2\theta}-rcos\theta \right )}
\end{eqnarray*}
\begin{equation}
\end{equation}
where the radial density distribution is a polytrope $\rho \propto r^{-n}$.

Since the cross section of Lyman $\alpha$ resonant scattering is a sharp peak at the core and  very extended on the wing,
 most scattered once photons are scattered on the far wing $x \sim 100 $ from the continuum of GRB's afterglow where $\tau \sim 1$. 
Eqs. (10) and (11) are good estimates when the time is small and the observed multiple scatterings are rare.
However, Eqs. (10) and (11) underestimate the intensity of photons near the emission peaks 
where optical depth is large. 
 Thus the scattered 
component in Figs. 2 and 3 are lower bounds. 
 Monte Carlo simulations in \S 3
 are able to find the true spectral profile and intensity.
Nevertheless, 
Eqs. (10) and (11) are accurate at small times and  give order of magnitude accuracy at later times. They are adequate to illustrate the formation  
of the Ly$\alpha$ emission feature. 
Accurate calculations in \S 3 will  push the predicted true emissions higher and make our conclusion stronger.

\begin{figure}%[htb]
\begin{center}
\includegraphics[height=6.5 cm]{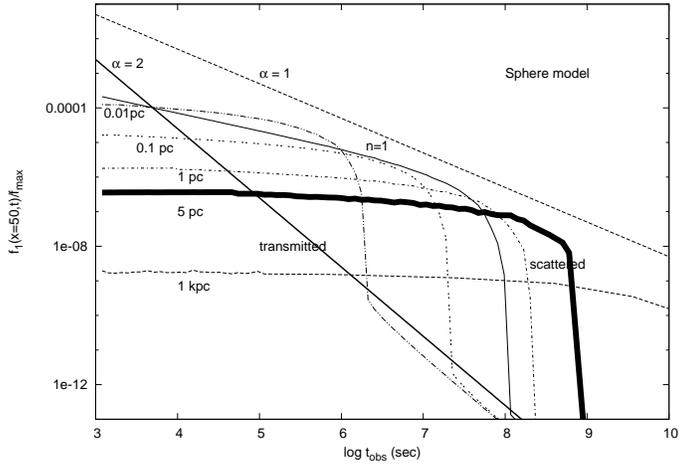}
\caption{Light curves of the scattered Ly$\alpha$ component at  $x=50$ (close to its peak frequency) for different cloud sizes in the Sphere model of circum-GRB neutral hydrogen. Also shown are  the transmitted component for photon source parameter $\alpha=2$ and $\alpha=1$, as well as the scattered component of a ploytrope distribution of neutral hydrogen (the thin solid curve with the label ``n=1" next to it). The fast decaying straight lines are for the transmitted component while the slow varing curves are for scattered component. Since the transmitted light decays fast, Ly$\alpha$ emission feature will be formed at times  where scattered light is brighter than the transmitted one. }
\label{fig2}
\end{center}
\end{figure}

\begin{figure}%[htb]
\begin{center}
\includegraphics[height=6.5cm]{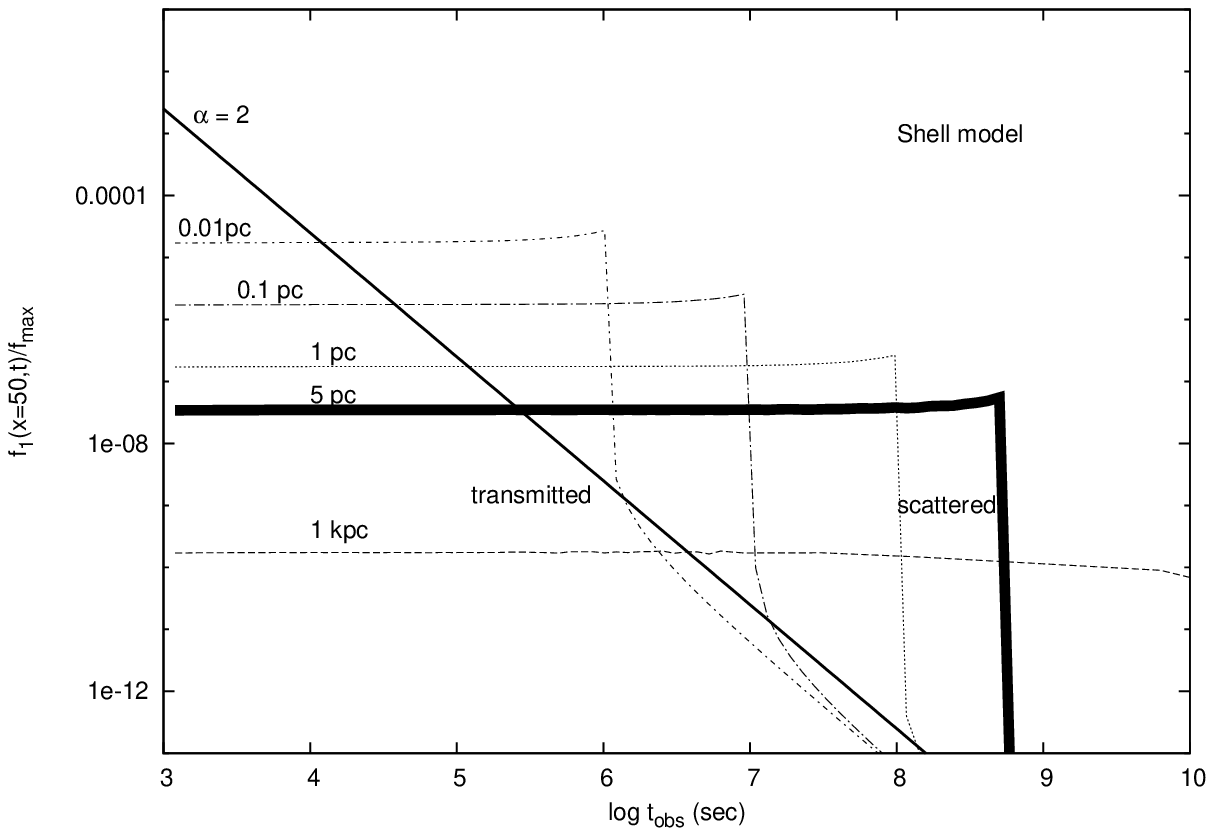}
\caption{As Fig.2 but for the Shell model.  The fast decaying line is the transmitted component while 
the slow varying curves are the scattered Ly$\alpha$ components
at $x=50$  for different cloud sizes in the Shell model of circum-GRB neutral hydrogen. Since the transmitted light decays fast, Ly$\alpha$ emission feature will be formed in the region where scattered light is higher than the transmitted one. }
\label{fig3}
\end{center}
\end{figure}

 In Fig. 2,  our simple analytic model gives a good idea on the formation of Lyman $\alpha$ emission and shows the light curves for a range of cloud sizes for the Shell model. 
The light curve of the scattered component is flat for a long time, in sharp contrast with 
the rapid decaying of the transmitted light. This is because  contributions from new 
areas are joining into the scattered component for the first time (Fig. 1b). The time
scale for the variance of the scattered component is the time scale of light crossing of the cloud.  When the cloud boundary is reached, we see a sudden drop of light by several orders of magnitude. After that, 
the scattered component becomes smaller than the transmitted component again.

 For our Shell model the luminosity of the scattered component is only about $10^{-7}$ of the maximum of the transmitted continuum. When the cloud size is  in the range of 0.01 pc to 1kpc, the range of the scattered component is between
$10^{-4}$ and $10^{-9}$. Adopting a different cloud column density does affect spectral shape but has little effects on the flux amplitude. This is because the peaks always correspond to $\tau=1$. The increased number of scattered photons is counteracted by more damping along their path of propagation to the observer.

  GRB progenitors are massive stars usually sitting in the middle of a 
density enhancement. If we assume a power law radial density profile of n=1 instead of n=0, our predicted brightness of the scattered component will 
increase by 3 orders of magnitude  at an observation time of one day (Figs. 2 and 3). This is because the intensity of scattered flux is sensitive to HI presence in GRB's immediate neighborhood, as a result of the time-delay pattern in Fig. 1b . A denser 
homogeneous cloud with $n\sim 100 cm^{-3}$ and radius $\sim$ 0.01 pc can produce similar  effects to a polytrope cloud 
of $n=1 cm^{-3}$ and radius $\sim$ 5 pc at times smaller than $10^6$ sec. (Fig. 2)

 It should be pointed out that the flatness of the light curve of the 
delayed arrival of scattered of $\tau \sim 1$
 photons, together with the rapid decaying of the source,  make the Lyman $\alpha$
 emission potentially identifiable. Should the source decays not fast enough ($\alpha=1$) ,
 the chance of telling the scattered emission from the transmitted one is  very slim. Also, if the cloud is too large ($r \sim 1$ kpc), the scattered emission may be too weak to be observed.

\section{Resonant scattering with Monte Carlo simulations}

\subsection{Method of Monte Carlo simulation}

Every new photon is released at the coordinate center along radial direction. The frequency distribution
of the new photon follows that of the continuum. Since continuum varies very little over a small frequency interval, we adopt a constant spectrum across the Lyman $\alpha$ profile. Once the photon enters the gas medium, the length of free path is determined by a distribution function $e^{-\frac{l}{l_*}}$ where $l_*$  is the mean length of free path. The location of the scattering is then determined. If it is outside the HI cloud, the photon is labeled escaped. 

At the new location of the scattering, the velocity $v=\frac{V}{V_D}$ of the HI atom is generated by two steps. First, the velocity components 
$v_x$ and $v_y$ ( $z$ is the propagation direction of photon) are generated following a Maxwell distribution $e^{-{v_x^2}}$. Second, the velocity $v_z$ is generated following the distribution:
\begin{equation}
f(v_z) \propto \frac{e^{v_z^2}}{(x-v_z)^2+a^2}
\end{equation}
which is the joint requirement of Gaussian distribution and Lorentz profile for the rest
frame cross section of resonant scattering. The distribution shown in equation (12) is not a true distribution
of velocities. From Eq.(3), $\Delta x=\vec{v}\cdot (\hat{e}_{photon'}-\hat{e}_{photon})$ when recoil is negligible. 
The velocity distribution in a scattering
thus represents  photon frequency
shifts with respect to the line center in velocity units. 
The direction of the resonantly scattered photon is assumed to be isotropic. Other distributions such as  dipole distribution (ZM02) would cause small differences. When a complete treatment with  polarization considered, the difference is limited to a factor of $25\%$ (Rybicki \& Loeb 1999). 
We restrict ourselves to isotropically scattered photons.
Once the direction is generated, frequency of outgoing photon can be calculated by Eq.(12). With this new set of frequency and direction of photon, we repeat the above procedures of calculating the free path and determining on the escape. Each photon is followed all the way along its path until it escapes.   

Since the effectiveness of generating $v_z$ determines crucially the speed of calculation, special algorithms have been proposed (Zheng \& Miralda-Escude 2002, ZM02 hereafter).
The distribution function of $v_z$ is a direct multiplication of two well known functions, namely Gaussian and Lorentz. They either have a sharp cut-off or a sharp peak. Since their probability peaks happen at different places, the total distribution function has two peaks. Because of this, any single step generation algorithm using rejection method will inevitably leave the peak of one function falling on the tail of the other function,  thus inefficient.  On the other hand, three or more step process will not improve the algorithm because the efficiency downgrades with more steps when computation overheads increase. As a result, ZM02's two steps algorithm works best. We also found that two step process may be an overkill when the two peaks are  close to each other ( for small $x$ at core scattering). We make an improvement on the algorithm and list it in the appendix. 50 million of photons are experimented for each model.

\subsection{A single blast of photons}

The observed flux of scattered photons
is a composite result of photons released at a series of epochs. At each epoch, the photon release is like a delta function.
In this section we study the photons released from a single moment. The source is strictly a $\delta$ function in time. We also assume that the source is at $z=0$. For such a photon source, the transmitted continuum flux exists only at the first moment (Eq.(9)). After that, resonantly scattered photons  start to arrive at the observer and become the source of the observed photons at times not equal to zero. We will show that the escaping time scale of these photons become the new time scale of these scattered emission.

Follow Eq.(10) and integrate the $\delta$ function over time for the source, the scattered-once light for the Shell model is 
\begin{equation}
f_1(x,t) = \frac{f_{max}}{2} \cdot 
\frac{\rm{arccos}(1-\frac{ct}{r})} {\sqrt{1-(1-\frac{ct}{r})^2}} \cdot \frac{c \tau_1}{r} \cdot \phi(x) e^{-\tau_{1} \phi(x)} 
\end{equation}
for $0\leq t \leq \frac{2r}{c} $ and is zero after that.
For small t, it can be approximated as
$
f_1(x,t) = \frac{f_{max}}{2} \cdot \frac{c \tau_1}{r} \ \cdot \phi(x) e^{-\tau_{1} \phi(x)} 
$.

Similarly, for the Sphere model
\begin{equation}
f_1(x,t) = \frac{f_{max}}{2} \cdot \int^R_0 \frac{dr}{R} \cdot \frac{c \tau_1}{r} \cdot
 \phi(x) e^{-\frac{\tau_{1} \phi(x)}{R} \cdot (r+\sqrt{R^2-r^2sin^2\theta}-r cos\theta)} 
\end{equation}

\begin{figure}%[htb]
\centering
\begin{center}
\includegraphics[width=6.0cm]{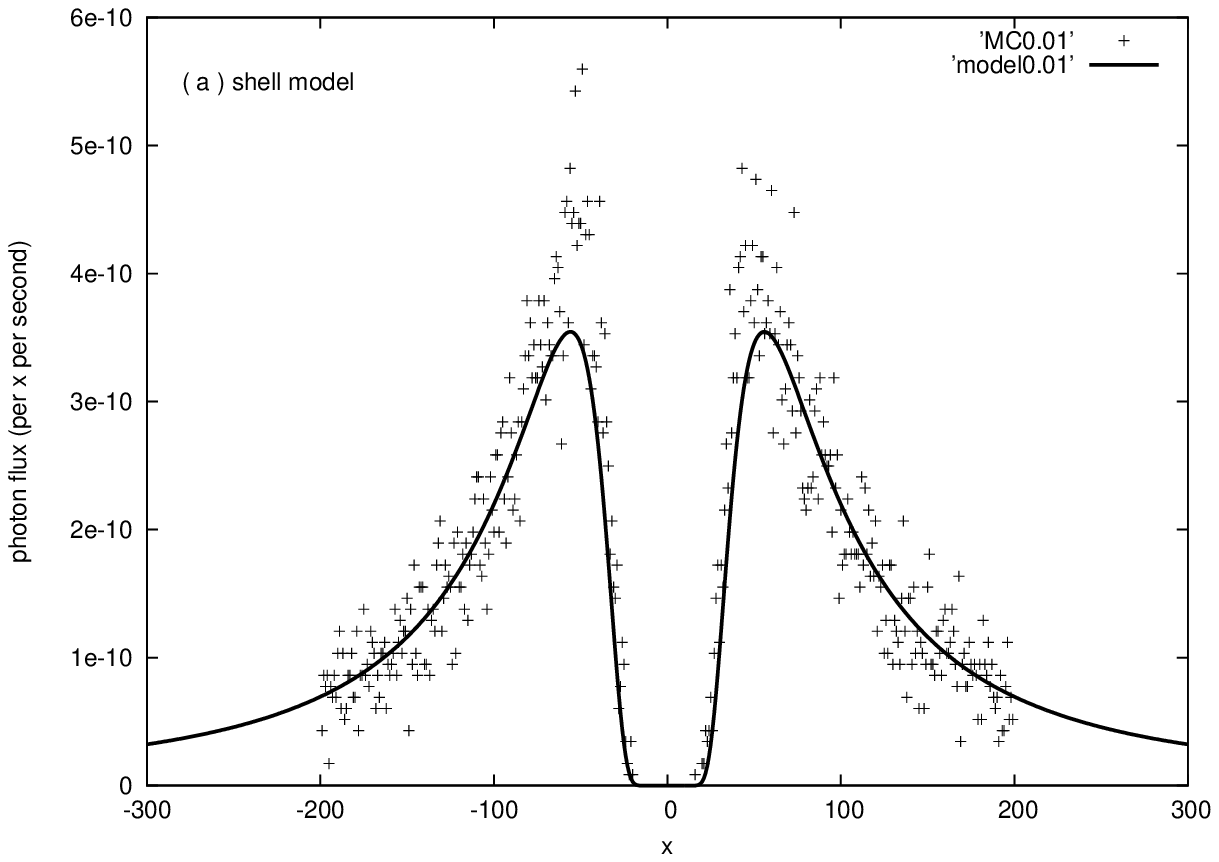}
\includegraphics[width=6.0cm]{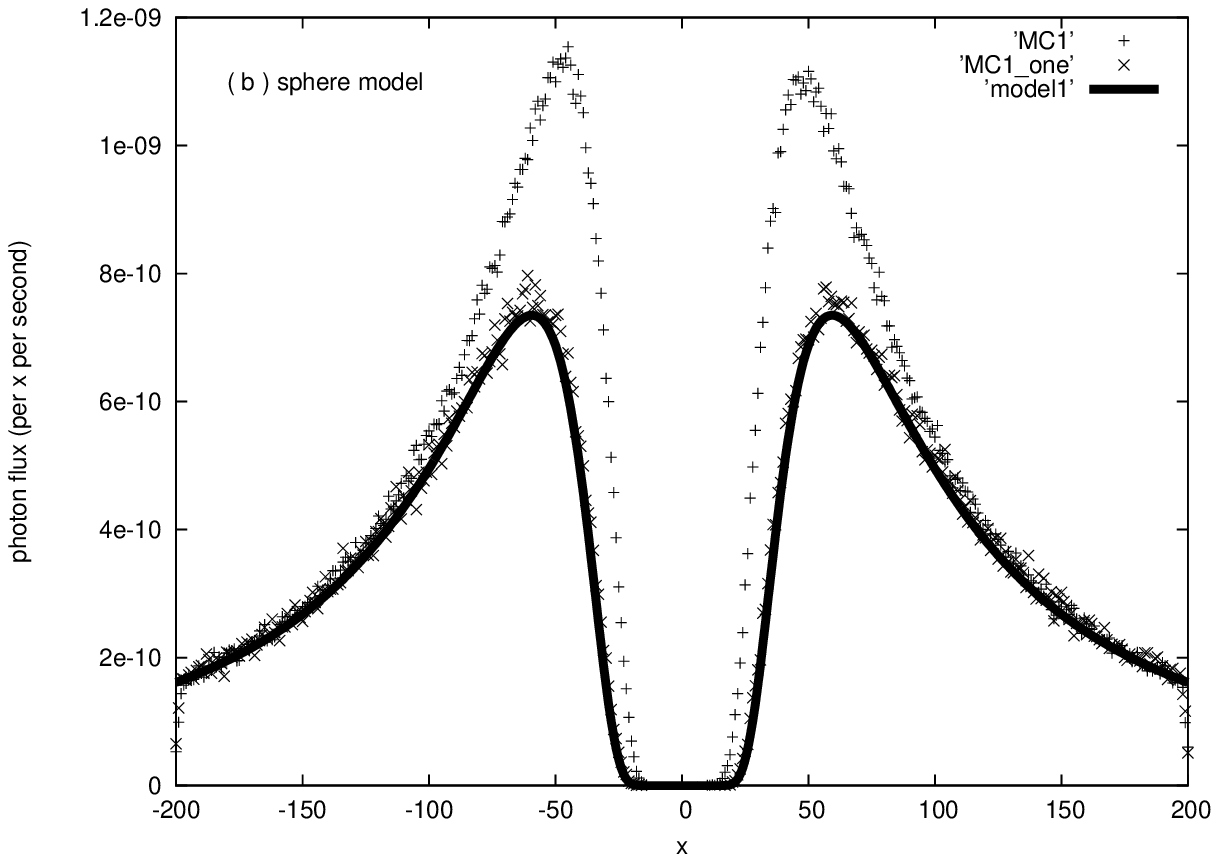}
\end{center}
\caption{ Emission flux derived from the analytic model is compared with  the results from MC simulation. 50 million photons are used in simulation. First graph is at $t=10^6$ sec for shell model of HI distribution, and the second graph is for the Sphere model at $t=10^8 sec$. The analytic model is a good approximation for $t < 10^6$ sec in both panels. The Sphere model result is a better prediction than that of the  Shell model at small times, but starts to have large errors for $t>10^8$ sec.}
\label{fig4}
\end{figure}

  The analytic results are compared with MC simulation results in Fig. 4.
For the Shell model, the simple analytic model is a good approximation before $10^6s$ in Fig.4a. The exception is at the central region where multiple scattered photons are hold responsible. At this early time, the location of the peaks are predicted correctly by the analytic model. This is because  90 percent of the escaped photons have scattered only once at this time (Fig. 5). For the Sphere model, similar good match is found at time smaller than $10^7s$.

\begin{figure}%[htb]
\begin{center}
\includegraphics[height=6.cm]{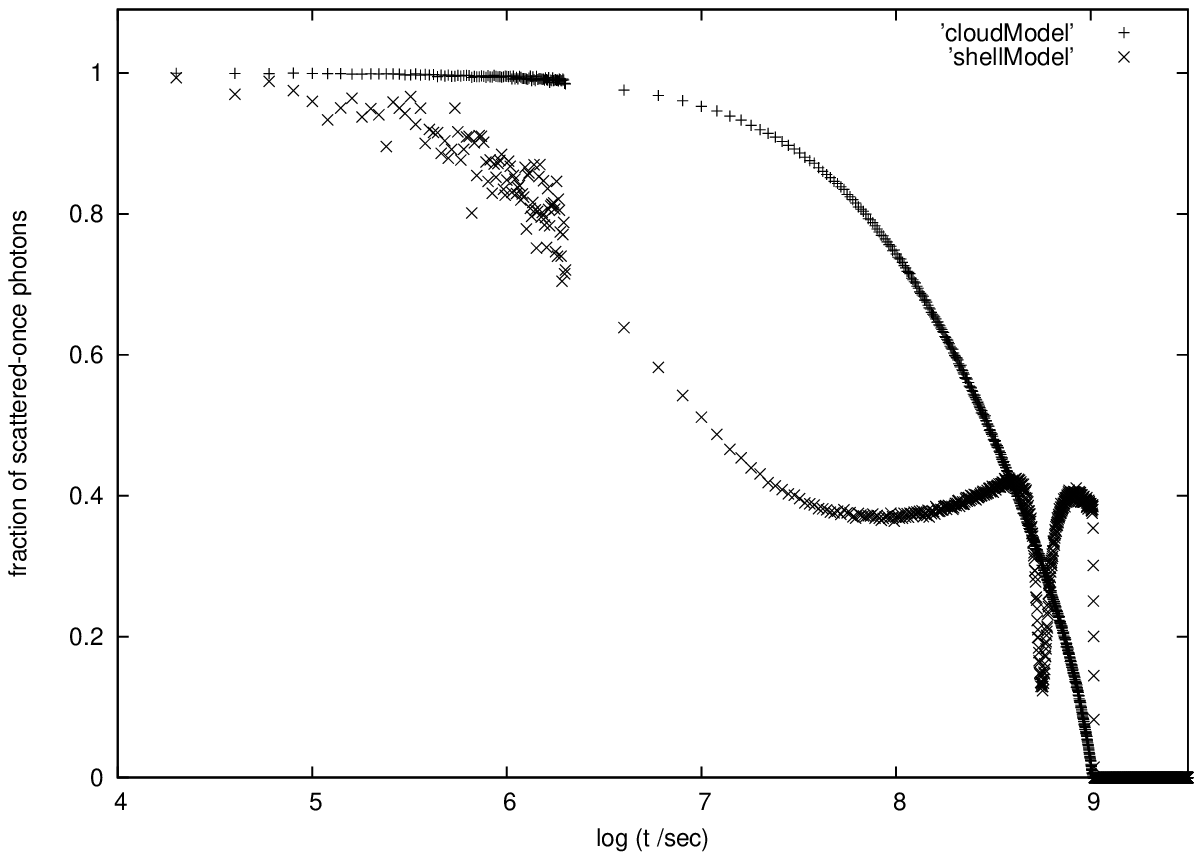}
\caption{ Evolution of percentage contribution of scattered once photons in the  scattered light. For the Sphere model, the scattered-once
photons are the major contributors of scattered component  until up to $10^8 sec$. For the Shell model,
photons which are scattered only once are more than $80\%$ of all the scattered photons for $t < 10^6$ sec. The analytic equation is not  a good
approximation  for $t > 10^7$ sec.}
\label{fig5}
\end{center}
\end{figure}

 In Fig. 4b, the scattered-once photons from MC simulation are sorted out and plotted separately. They are found to be in good agreement with the analytic model predictions and are the major contributors of the photon flux on the wing. But at $10^8 s$ for the Sphere model ($10^7s$ for the Shell model), the simple analytic model begins significantly underestimating the emission flux.  This is because more than 50 percent of the escaped photons are now scattered more than once (Fig. 5). For the Shell model, such time is at $10^7 s$.

 For large t $>10^9s$, MC simulations start to give different results from analytic approximations.
For both the Shell model and the Sphere model, the separation of the peaks from MC grows smaller with time,  
and the peaks become much higher than the analytic model predictions. The profiles are very different from those of analytic model, too. The MC profiles are fatter at the center. The analytic model has significantly underestimated the flux  at the center frequencies. MC results find photons at frequencies where the
analytic model predicts nil flux. These are the photons from multiple scatterings (Fig. 4b and Fig. 5). The analytic model places a lower bound and is a poor fit to the emission profile at late times.

\begin{figure}[htb]
\centering
\begin{center}
\includegraphics[width=6.0cm]{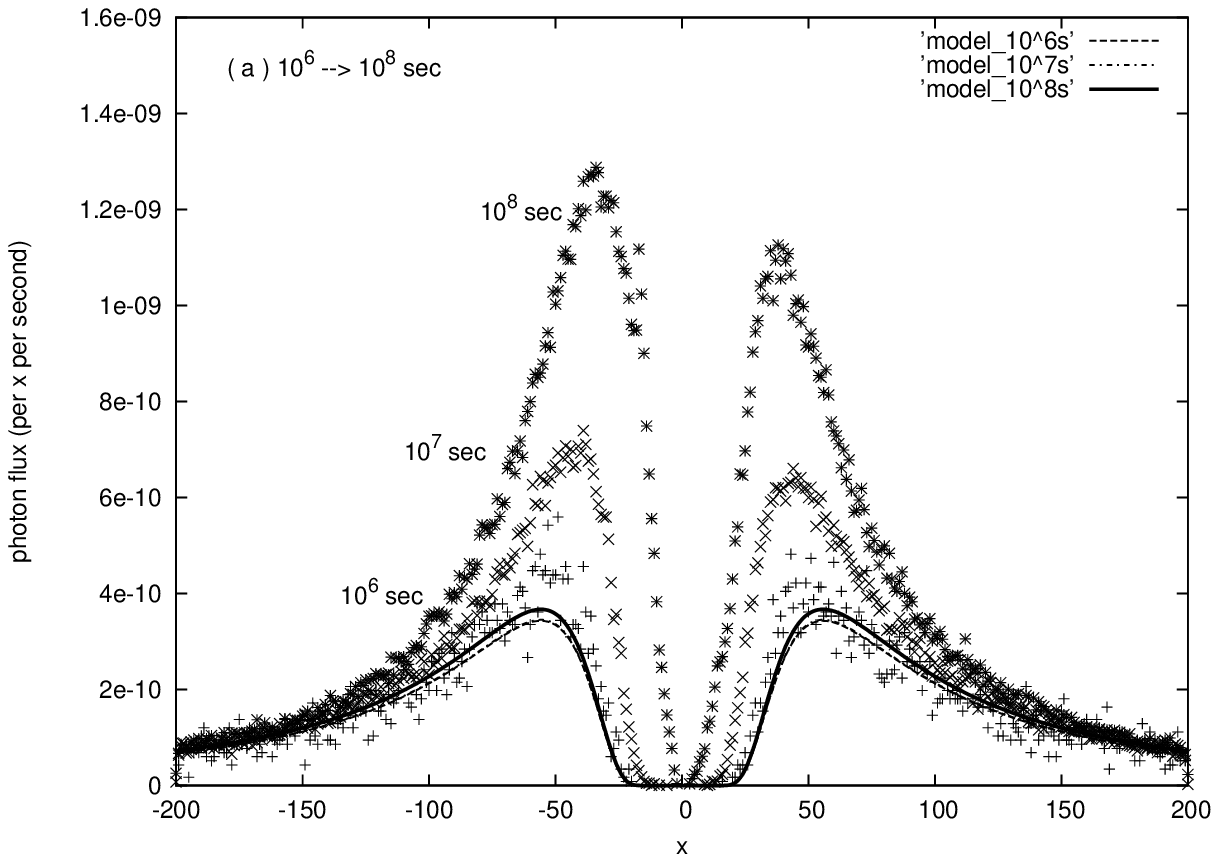}
\includegraphics[width=6.0cm]{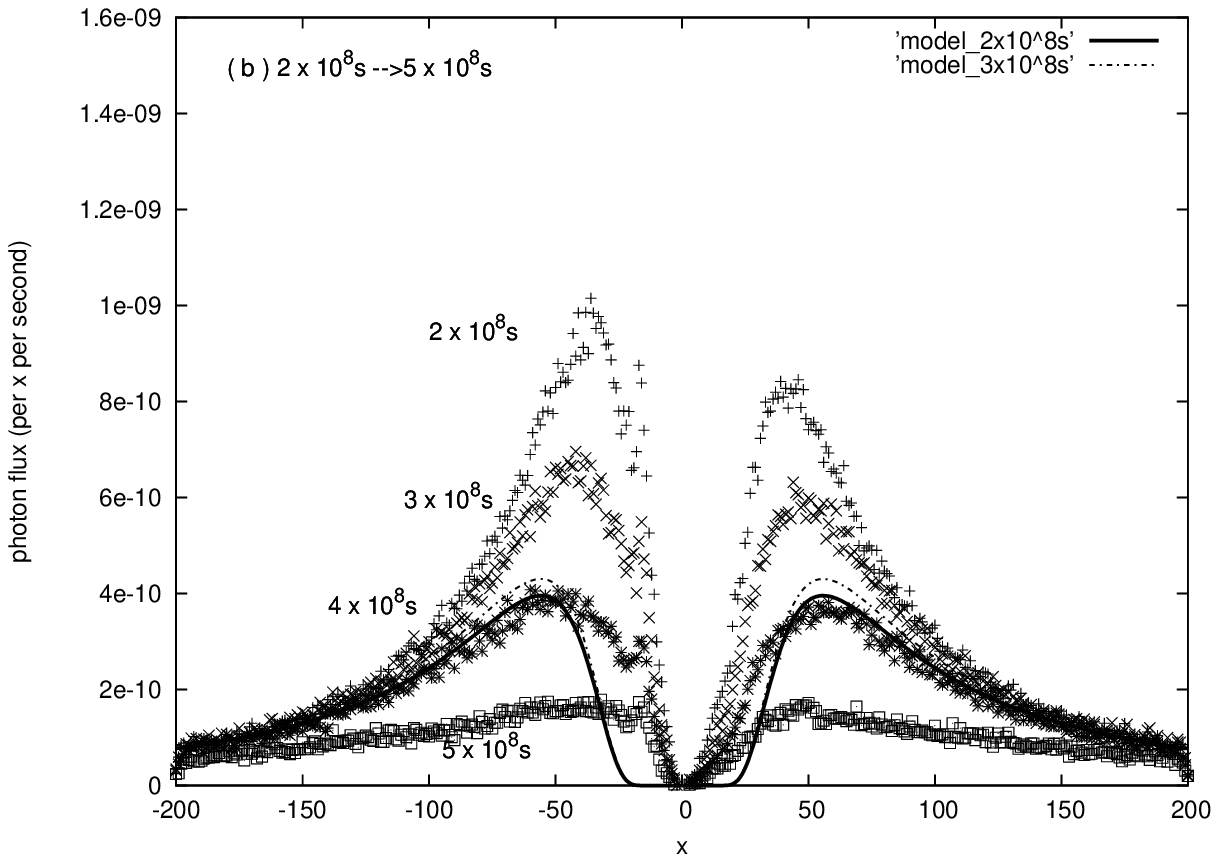}
\end{center}
\begin{center}
\includegraphics[width=6.0cm]{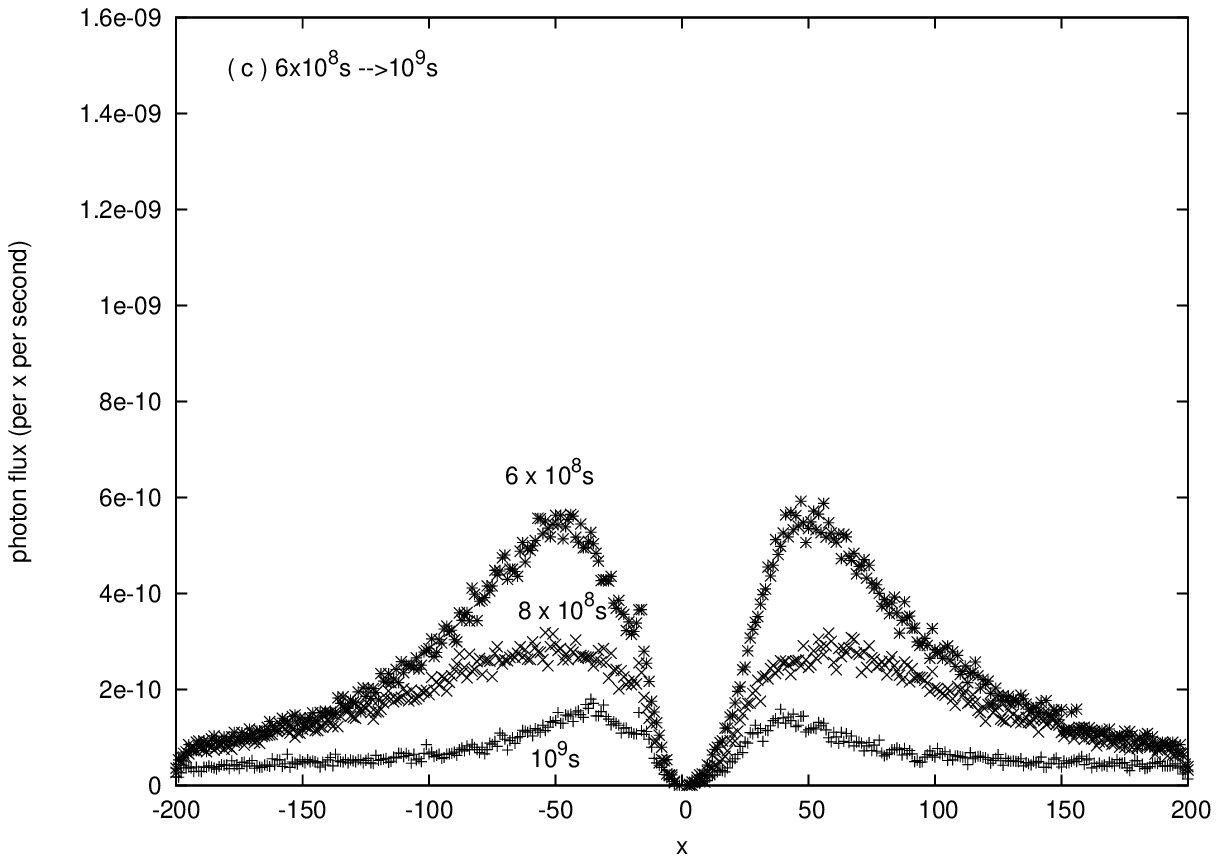}
\includegraphics[width=6.0cm]{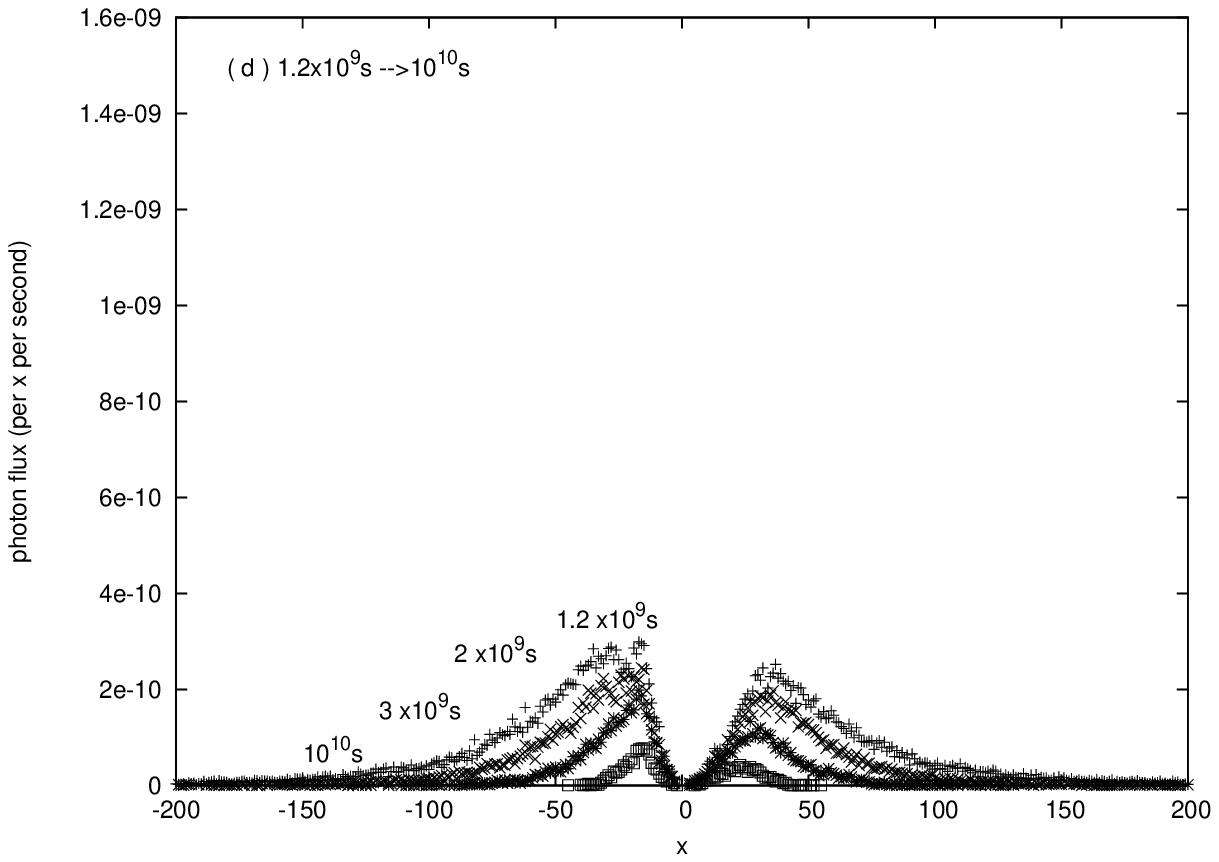}
\end{center}
%\vspace{4cm}
\caption{ Evolution of Lyman $\alpha$ emissions formed by scattered photons from a flash of continuum light in the Shell model. Flux is in unit of  photons per x per second. Data points are from MC simulations at different escape epochs. In Panel a, the emission intensity increases from epoch  $10^6$ to $10^8$ sec. In Panel b, the emissions decrease with time  from $2\times10^8$
to $5\times 10^8$ sec, and  from $6\times10^8$
to $10^9$ sec in Panel c. The emissions increase from $10^9$ to $1.2\times10^9$ sec when the scattered photons from the far side of the shell
 arrive. In Panel d, emissions decrease monotonically with time. Some analytic model results are drawn  with lines in Panels a and b at  $0.01, 0.1, 1, 2, 3 \times 10^8 $ sec. }
 \label{fig6}
\end{figure}

Fig.5 shows the percentage of the scattered-once photons in the total flux of the scattered light in the frequency range $x \in [-200,200]$.
The second peak at a large time for the Shell model comes from the scatterings at the far side of the shell.
  At times $t< 3 \times 10^5 $sec for the Shell model and $t<2 \times 10^7 $sec for the Sphere model, more than $80 \%$ photons are scattered only once. 
Compared to the Shell model, the Sphere  model has higher contribution of scattered-once photons and are better approximated by our analytic model.
This is because the neutral gas is more uniformly distributed in the Sphere model. A sphere may be thought of as a group of shells. Continuously there was a new addition of source regions into the contribution of
scattered-once photons.   As a result, the once-scattered domination regime lasts longer in the Sphere model.

Fig.6 shows the evolution of resonantly scattered photons for the Shell model. The MC simulation results are measured at observation epochs between 
$10^6$ sec to  $ 10^{10}$ sec  for a flashed
release  of continuum photons.  In Panels a and b, results for both MC simulation and analytic model are 
drawn. The 
analytic approximation does not predict the correct shape of the profile but can be treated as the asymptote at large x (Panels a \& b). The Lyman
$\alpha$ emission grows with time before $t < 3 \times 10^8 $ sec (Panel a) when the spherical angle of the first scattering region becomes larger 
(Eq.(10)), and then decreases with time till $5\times 10^8$ sec (Panel b) when the spherical angle of the first scattering region reaches $\theta=\frac{\pi}{2}$. After that it increases with time again till 
about $7 \times 10^8$ sec (Panel c) when the scattered-once photons from the far side of the shell arrives, and then decreases with time monotonically afterwards(Panel d). At later times, the luminosity decays and the peaks become closer to the center and the profile becomes narrower.

\begin{figure}%[htb]
\begin{center}
\includegraphics[height=6.cm]{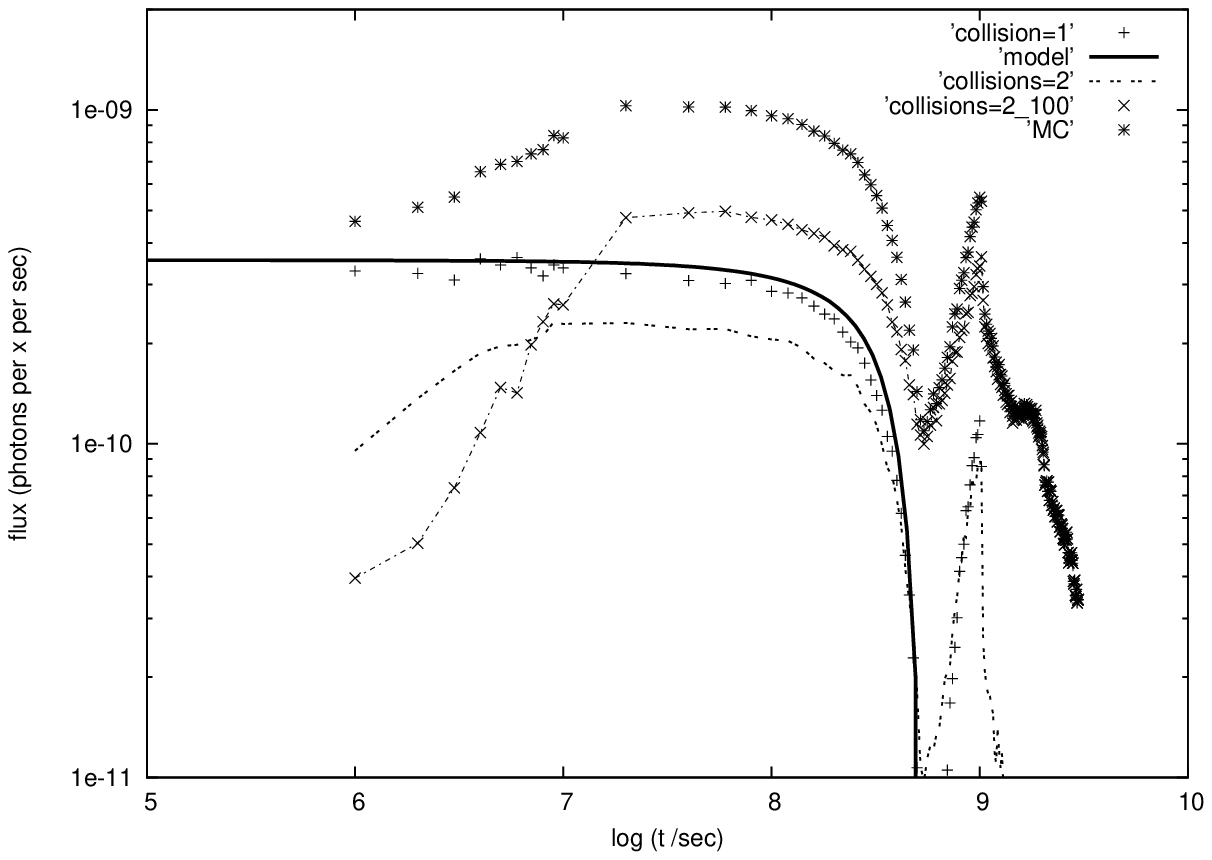}
\caption{Flux of photons as a function of time. The heavy crosses at the top are MC results for the Lyman emission. The individual contributions from photons which experience a total  scattering number of one, two, 2 to 100 times are also marked. The solid line is our analytic model Eq.(10) which includes scattered-once photons. The second peak is caused by resonantly scattered photons in the gas from the far side of the GRB.  }
\label{fig7}
\end{center}
\end{figure}

Fig. 7 shows the evolution of photon flux at  frequency offset $x=50$ 
 for a single flash of photons released at the source in the Shell model of neutral hydrogen.  Fluxes from MC simulations are
grouped with $10^6 $ sec interval. The contributions of photons experiencing a specified number of total collisions 
are shown in the figure for the flux at frequency $x=50$ . It is evident that the scattered-once photons dominate the total Lyman $\alpha$ emission before $10^6$ sec. Then, the scattered twice photons begin to dominate till $10^7$ sec. And from $10^7 $sec on, the 
Lyman $\alpha$ emissions mainly consist of photons which experience more than 2 collisions but less than 100 collisions. Photons of higher number of collisions has a flux lower than $10^{-11}$ 
(photons per $x$ per second), making a negligible contribution to Ly$\alpha$ emission for the whole duration of our interest. This nil contribution are the result of two facts. First, the source of the scattered photons are the continuum photons. Most scattered photons come originally from a wing frequency at which the cloud is optical thin ($\tau \sim 1 $ at $x \sim 100$). Multiple scatterings mostly happen at the core frequencies $x\sim 1$ which 
 take only a tiny section of the continuum ($x\sim 1$). Second,   photons which are scattered millions of times are spreaded
 over a longer period to escape.  
Fig. 7 shows good agreement about scattered-once photons between MC's result and that of the analytic model. Before $5\times10^8$ sec when half of the whole spherical shell is reached, Eq.(10) is a valid prescription.   

\begin{figure}%[htb]
\begin{center}
\includegraphics[height=6.cm]{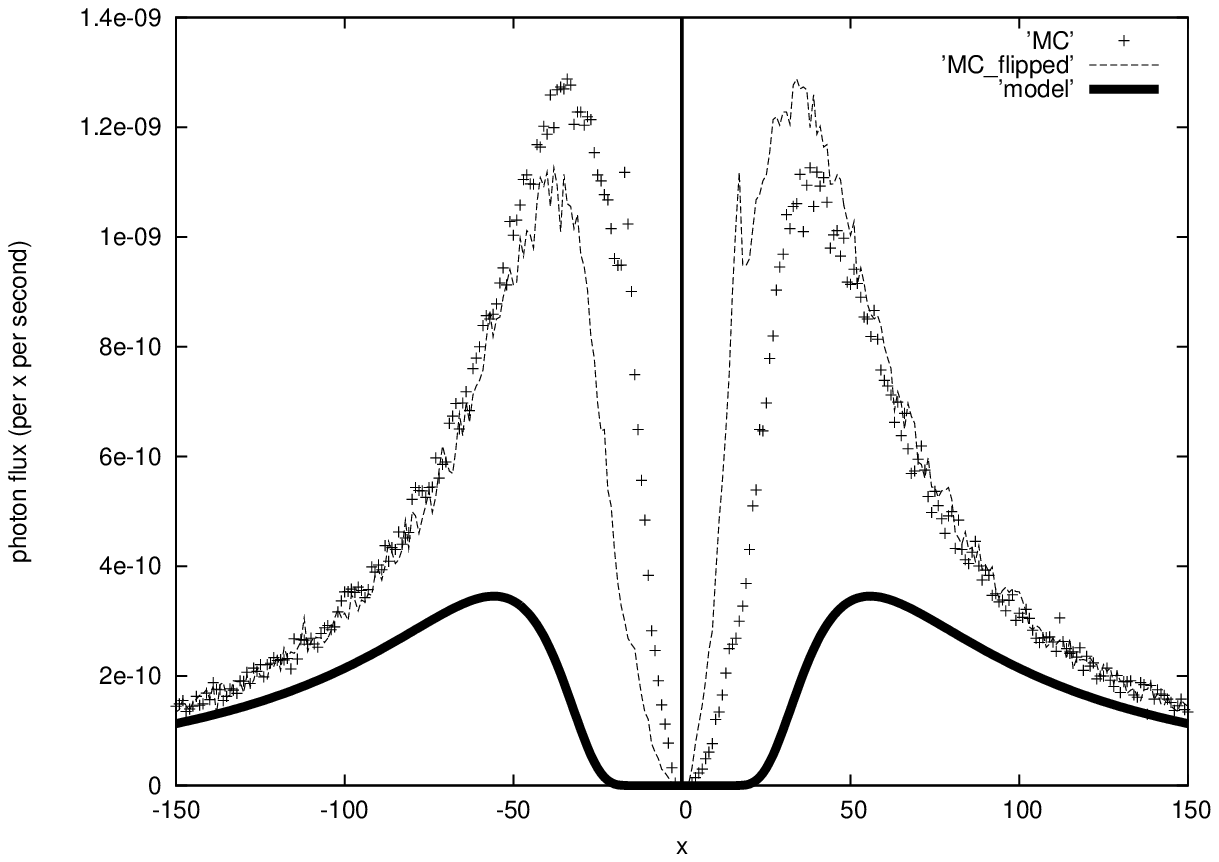}
\caption{ Asymmetry of red and blue peaks. Shown are the scattered emission profile for the Sphere model at $10^8$ sec from both the MC simulation and our analytic model Eq.(10). To guide the eyes for comparison, we have flipped the MC simulation profile left and right and drawn with a dashed line.  A center line is also added. }
\label{fig8}
\end{center}
\end{figure}

If recoil is ignored, the scattered peaks will be twin peaks exactly symmetric about $x=0$ where the line center of Lyman $\alpha$ is. The analytic model always predicts symmetric profiles, and the MC results at early moments are symmetric, too. However, spectral profiles at later time become slightly asymmetric. The red peaks  will get more photons as a result of recoil. This asymmetry is a reflection on the asymmetry of the number of red and blue photons which are scattered multiple times 
in the cloud. This is the so-called  Wouthuysen-Field effect as discussed by many authors (Wouthuysen 1952; Field 1958, 1959; Roy et al 2009c).
Fig. 8 shows the symmetry comparison of the spectral profile. Asymmetry exists but is  small.
 The red peak seems to be able to completely include the blue peak.

MC simulations show strong profile evolution of the scattered
 Lyman $\alpha$ emission. This evolution can be best characterized by the change in the frequency
location of the emission peaks. In Fig.9, MC simulation results evolve from a larger separation of peaks to a smaller value of Adams (1972)'s prediction. Our analytic model predicts the correct initial value but fails at larger times, especially for the Sphere model (Fig.9b).

\begin{figure}[htb]
\centering
\begin{center}
\includegraphics[width=6.0cm]{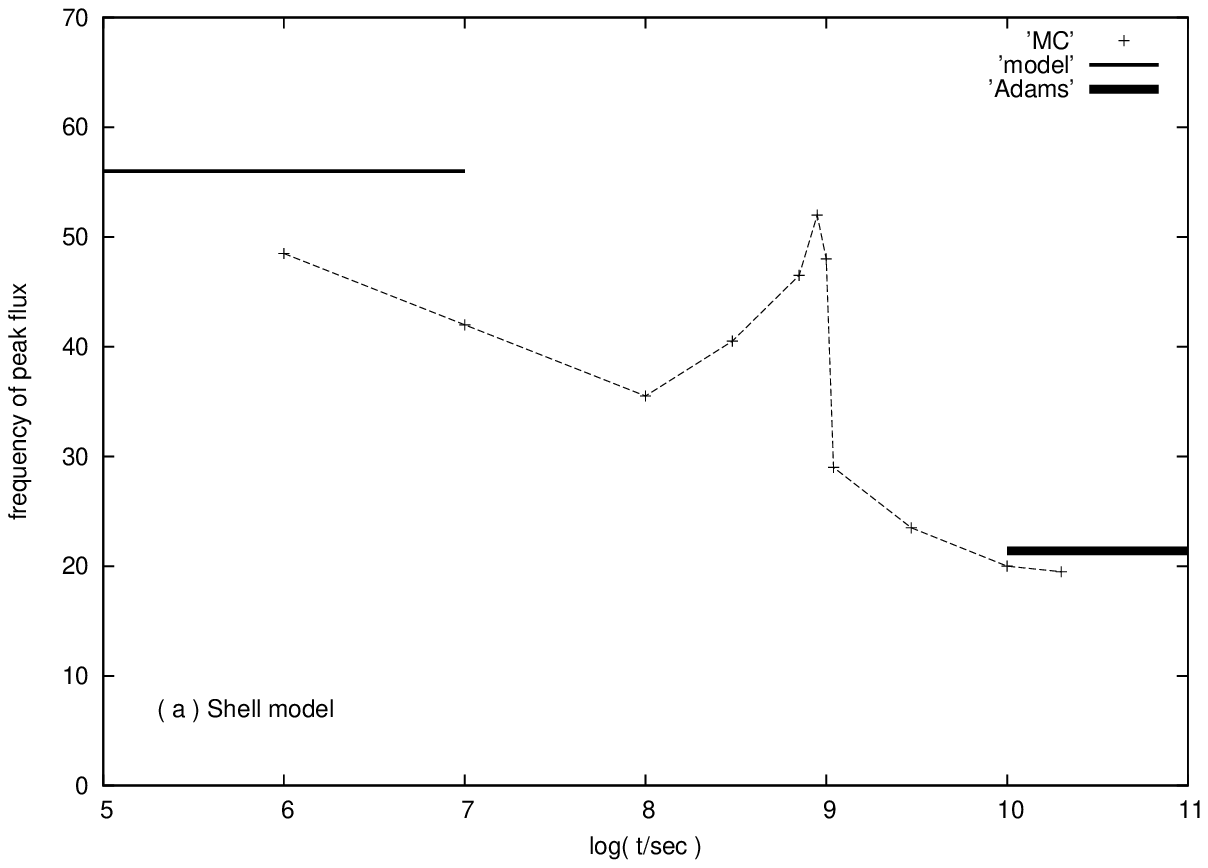}
\includegraphics[width=6.0cm]{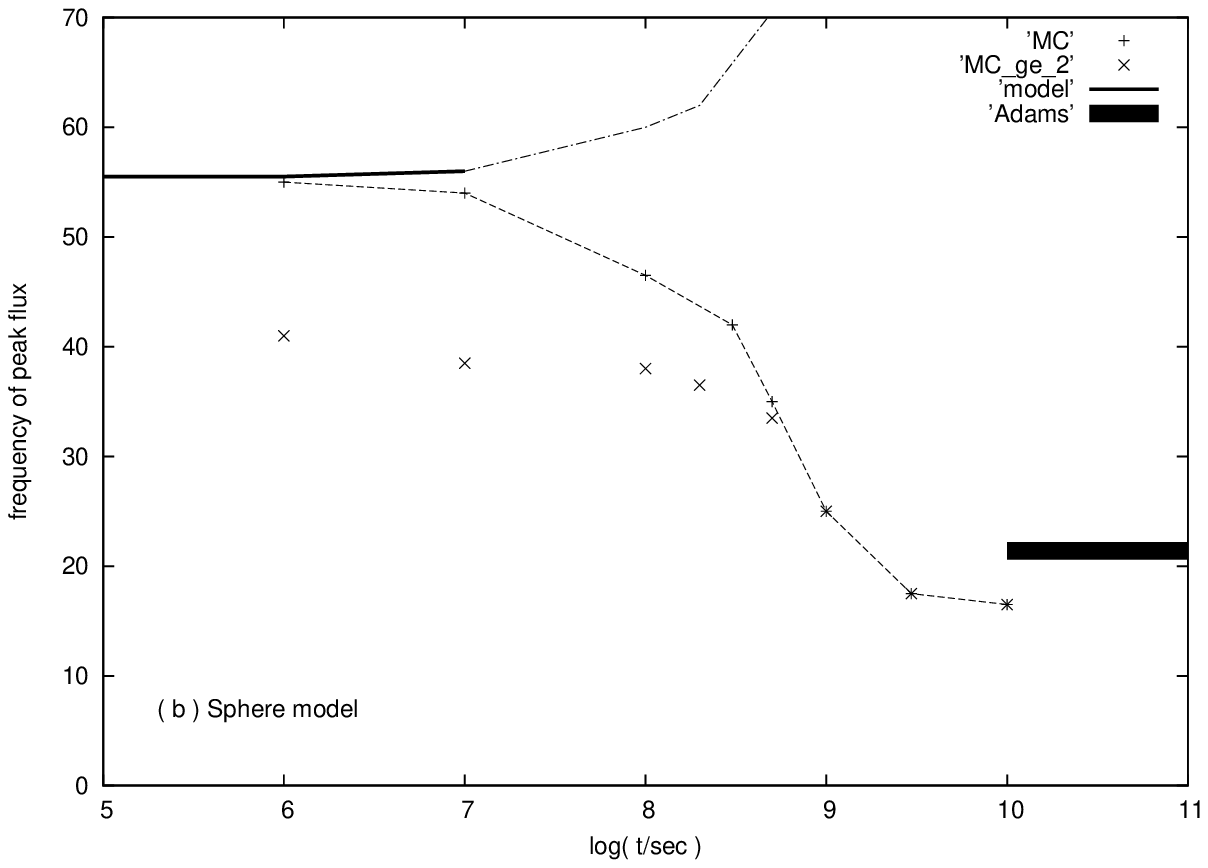}
\end{center}
%\vspace{4cm}
\caption{ Evolution of the frequency offset of the peak fluxes of the emergent scattered photons, (a) for the Shell model, (b) for the Sphere model. The MC simulation results  of the blue and red peaks are averaged and plotted as connected points. The model predictions at small time (our Eq. (15)) and at late time (Adams 1972) are shown with thick solid lines. In panel (a) the peak around $10^9 $ sec is caused by photons scattered from the far side of the HI shell around GRB. In panel (b) our analytic model in \S 3 gives poor predictions when time is larger than $10^7$ sec, thus marked as dashed line from there. At a later time,  MC results approach Adams' prediction but are different by a small factor.}
 \label{fig9}
\end{figure}

From Eq. (10), the flux
$ f_1 \propto \phi(x)e^{-\tau_{1}\phi(x)}$ for the Shell model. This relation is also a good approximation
 for the Sphere model at small times. 
The relation is
 $f_0 \sim  \phi(x) \sim \frac{1}{\Delta \lambda^2}$ on the red side, and 
$f_0 \sim e^{-\tau_{1}\phi(x)}$ on the
blue side, which is steeper than an exponential cut off. By taking derivative,  
the peak is at $\tau_{1}\phi(x_{peak})=1$ so that 
\begin{equation}
x_{peak}=\left(\frac{a\tau_1}{\pi^{\frac{3}{2}}}\right )^{\frac{1}{2}}, or,  
\Delta \lambda_{peak} \approx {1.45\AA} \sqrt{\frac{N_{HI}}{10^{20.3} cm^{-2}} }
\end{equation}
This is different from the peak position at later times which was predicted (Adams 1972) as 
$x_{Adams}=\left ( \frac{a\tau_1}{\pi} \right )^{\frac{1}{3}}$ (
slightly different from its original form because 
in our notation the central optical depth is $\tau(0)=\tau_1 /\sqrt{\pi}$).
Our peak position is related to Adams' by 
$x_{peak}= x_{Adams} \cdot (a\tau_1)^{\frac{1}{6}}\pi^{-\frac{5}{12}}=2.85 \cdot x_{Adams}$ for
our assumed gas column density and Doppler velocity.  The discrepancy shows that the peak location is an evolving quantity. The two different  predictions correspond to limiting cases at different times. Our analytic Eq.(15) is accurate at times immediately after the burst, while Adams' prediction describes behavior at  very large times. 

Profile evolution is characteristic of the Lyman $\alpha$ emission formed by our mechanism. As time passes on, the profile becomes narrower and more centrally peaked because photons which are scattered numerous times
 become the major component of the emergent emission. Fig.9 shows such spectral evolution.
The frequency offset of the peak fluxes are drawn for the emergent scattered photons. We see a clear declining trend. The value starts well from our simple model prediction.  At a later time when photons experiencing multiple collisions simply add on top of a slowly varying profile of scatter-once photons, the
peak locations are determined by these multiple scattered photons. Fig. 9 shows how peaks change their locations while their constituents change from scattered-once photons into photons experiencing multiple scattering as time increases.  Our analytic model predicts the initial values where all peaks start with, while Adams'(1972) result corresponds to the asymptotic final value of the peak location at large t. Our MC results at large time agree with Adams' prediction but are smaller by a minor numerical  difference. On the other hand, in Adam (1972)'s example case of plane parallel atmosphere, their numerical result is slightly higher than their analytic prediction. In general, a fractional difference is expected to reflect the difference in  scattering geometry. We see small difference between our two models of HI distribution.

\begin{figure}%[htb]
\centering
\begin{center}
\includegraphics[width=6.0cm]{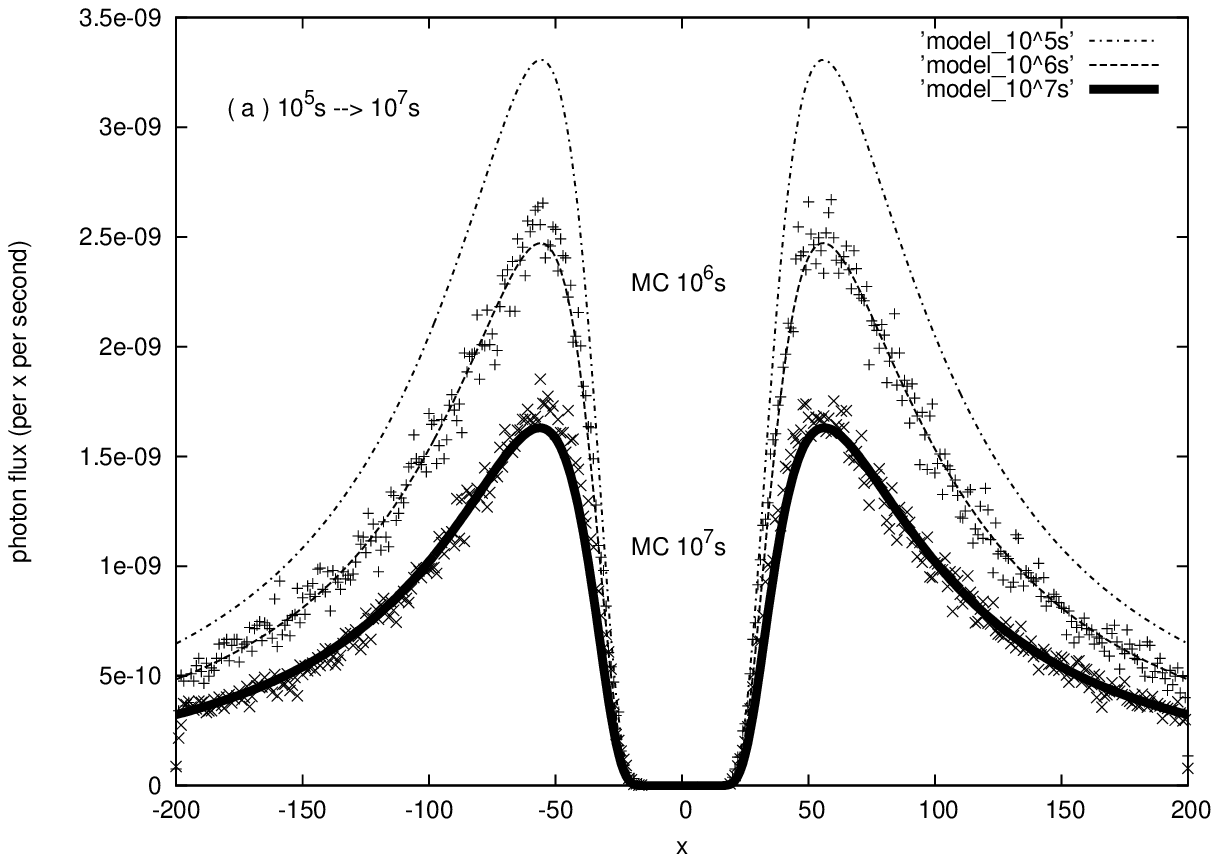}
\includegraphics[width=6.0cm]{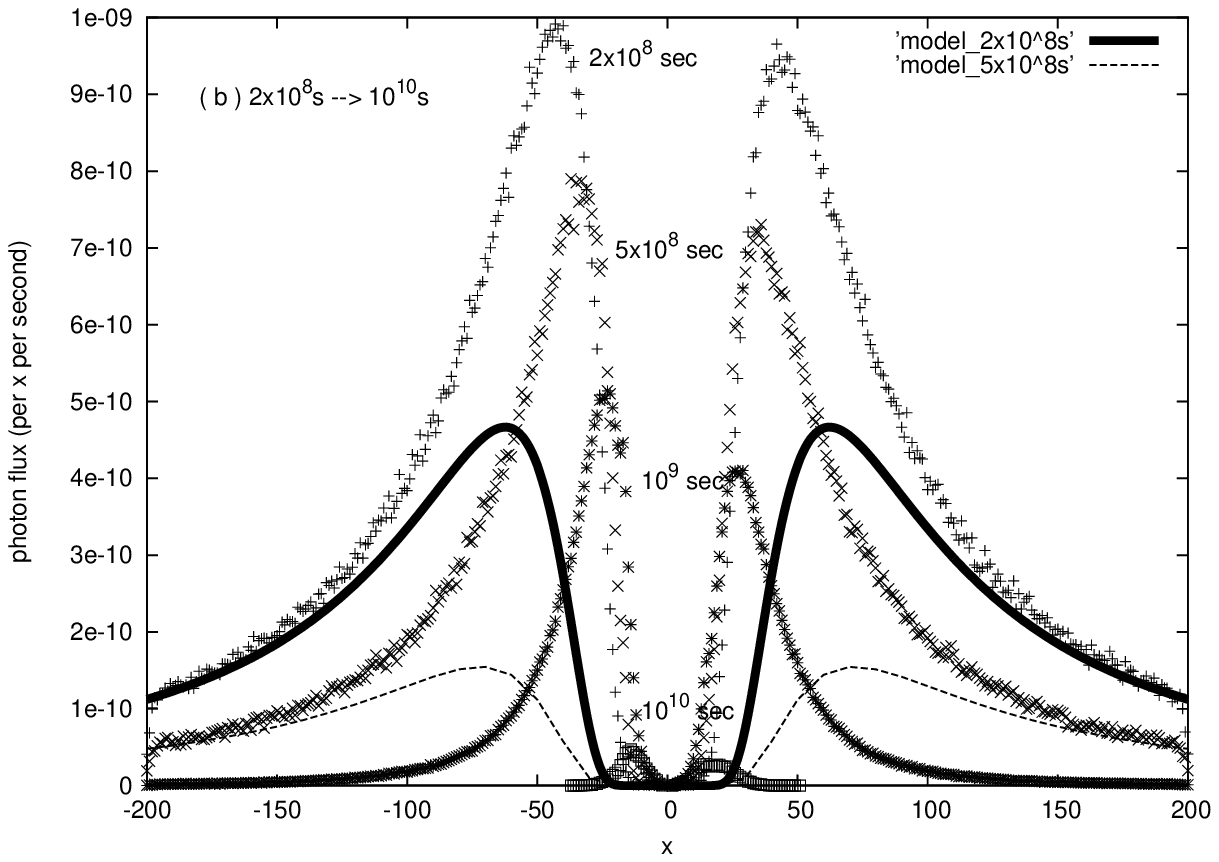}
\end{center}
\caption{ Evolution of resonantly scattered photons as measured at various emergent epochs for a flash of continuum source in the Sphere model. Flux is in unit of  photon per $x$ per second.  Monte Carlo simulation results are shown  at $10^5$ and $10^8 $ sec in panel (a), at $2, 5, 10, 100\times 10^8 $ sec in panel (b). Lines represent the analytic model  results at $0.001, 0.01, 0.1, 2, 5 \times 10^8 $ sec.
Intensities decrease with time in the figure.
 In panel (a),  both MC simulation and analytic model give similar results for times smaller than $10^7$ sec. In panel (b) at later times, the luminosity decreases and the peaks become  closer and narrower with time.  }
\label{fig10}
\end{figure}

In Fig. 10 shows how Lyman $\alpha$ emissions evolve with time in the Sphere model of HI distribution. In Panel a, the analytic model agrees well with MC simulations at small times $10^6$ and $10^7$ sec. Unlike the Shell model (Fig.6), the luminosity of Sphere model decreases with time monotonically. 
At larger times, the analytic model starts to significantly underpredict the luminosity (Panel b), especially at frequencies closer to the line center. Also, the analytic model fails to predict the evolution of profile which is evident at later times. From MC simulations, the emission peaks are found to shrink monotonically in amplitude, frequency location and  the width of peak.

\begin{figure}%[htb]
\centering
\begin{center}
\includegraphics[width=6.0cm]{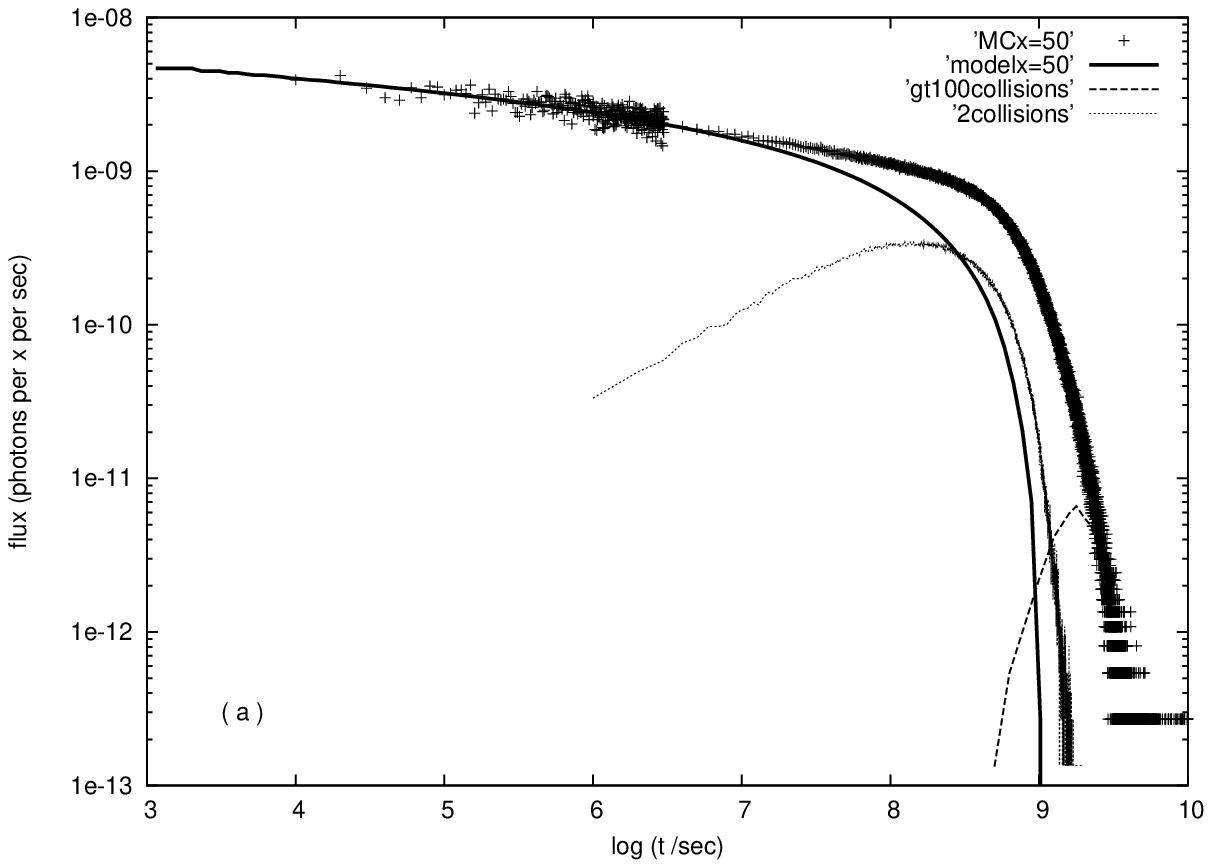}
\includegraphics[width=6.0cm]{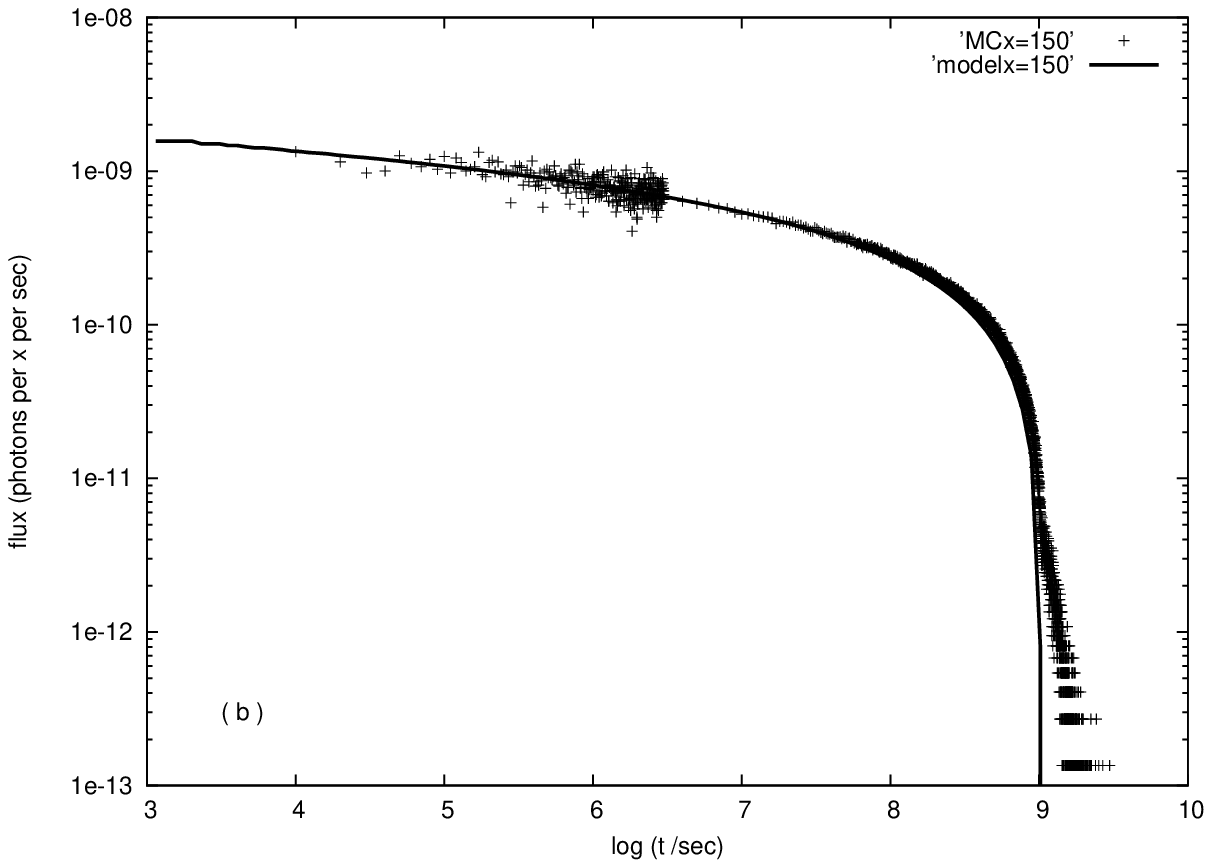}
\end{center}

\caption{ Evolution of photon flux at two different frequencies (a) $x=50$  and (b) $x=150$  for a single flash of photon source in the Sphere model of neutral hydrogen. The solid line is the analytic model prediction which represent photons of scattering once. Fluxes from MC simulations are shown as marked points, which are grouped with interval $10^6 $sec after $t=3\times10^6$ sec and $10^4 $sec before that. The contributions from photons which are scattered  twice, and greater than 100 times from MC simulations are individually shown in panel (a) as dashed lines. }
\label{fig11}
\end{figure}

Fig.11 can  be thought as the distribution of the delayed arrival time of the scattered photons from MC simulation
for a single flash in the Sphere model. 
The scattered once photons cease completely at about $10^9$ sec, corresponding to the round trip travel time  for a photon to pass straightly from the center to the cloud's opposite surface and then bounce back to get through the entire sphere.
This is the characteristic light crossing time in the scattering geometry.
 In Fig. 11a, for observation time of interest ($<10$ years),  scattered once photons are the major contributors.
Even for photons which are scattered exactly twice, their contribution to the total photon flux is negligible until at later times when scattered once photons become rare (Fig.5). 
 At late times, the photon flux contribution seems to consist of photons with multiple scattering history.

For scatterings at the far wings (Fig. 11b, $x=150$), it would be very rare for a scattered once photon to get a second scattering before it escapes.
This is why the simple analytic model prediction agrees so well with the MC simulations.  Multiple scattered photons show their traces 
only when the scattered once photons completely cease to appear. The next major contributors are found to be the scattered twice photons. 

\begin{figure}%[htb]
\begin{center}
\includegraphics[height=6.cm]{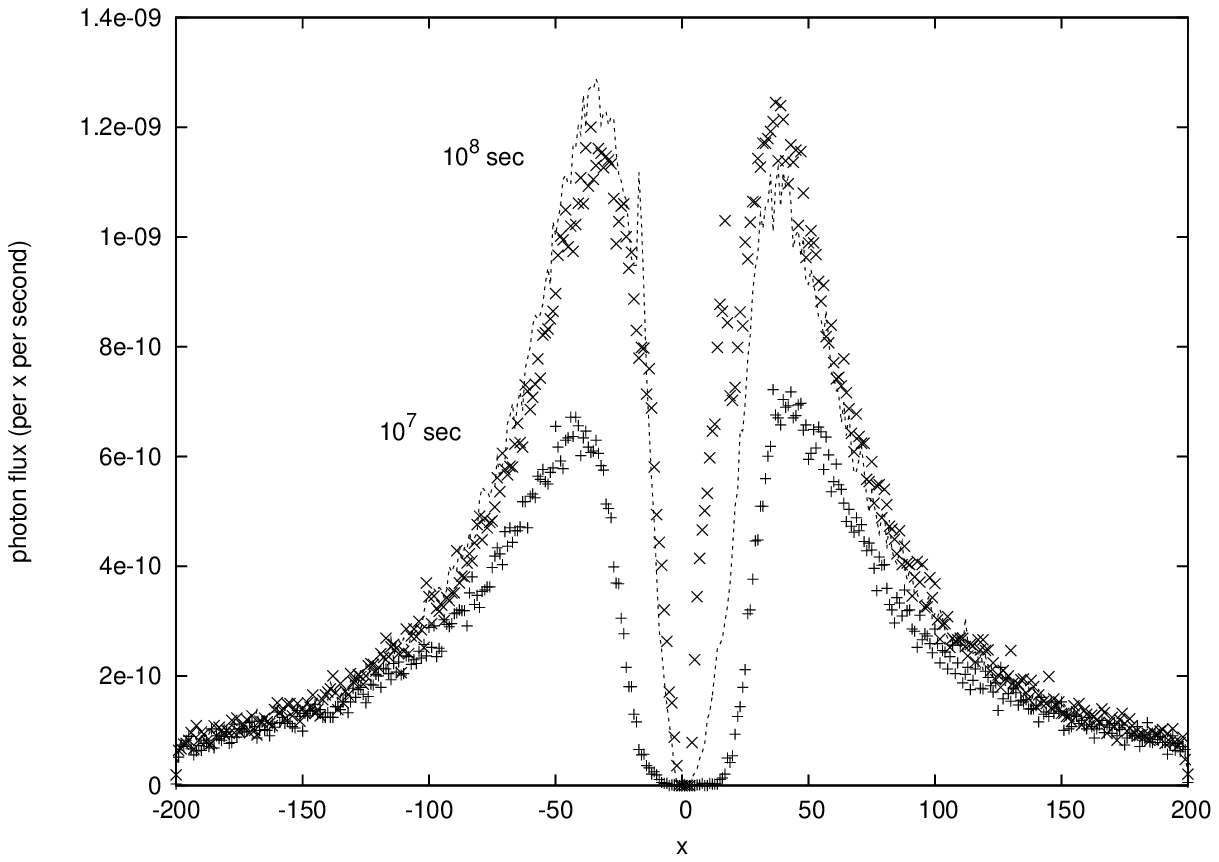}
\caption{ The effects of shell's moving on the emission profile in the Shell model.  MC simulation results at $10^8$ sec and $10^7$ sec 
are shown for an inward moving shell with constant speed $V_D$.
The results of a static shell (Fig.6a) at $10^8$ sec is drawn as the dashed line for comparison. }
\label{fig12}
\end{center}
\end{figure}

So far we have assumed HI has no bulk motion relative to the GRB. In reality, gravitation or hydrodynamic feedbacks of soft X-ray and UV radiation of either GRB or its progenitor may accelerate the gas. In Fig.12 we discuss the Shell model when motion of the HI is considered. We have assumed a constant inward speed of $V_D$ for all the HI on the shell. Spherical bulk motions may change the symmetry of the spectral profile.

\subsection{Synthesized MC simulation results for  GRB's afterglow}

\begin{figure}%[htb]
\begin{center}
\includegraphics[height=6.cm]{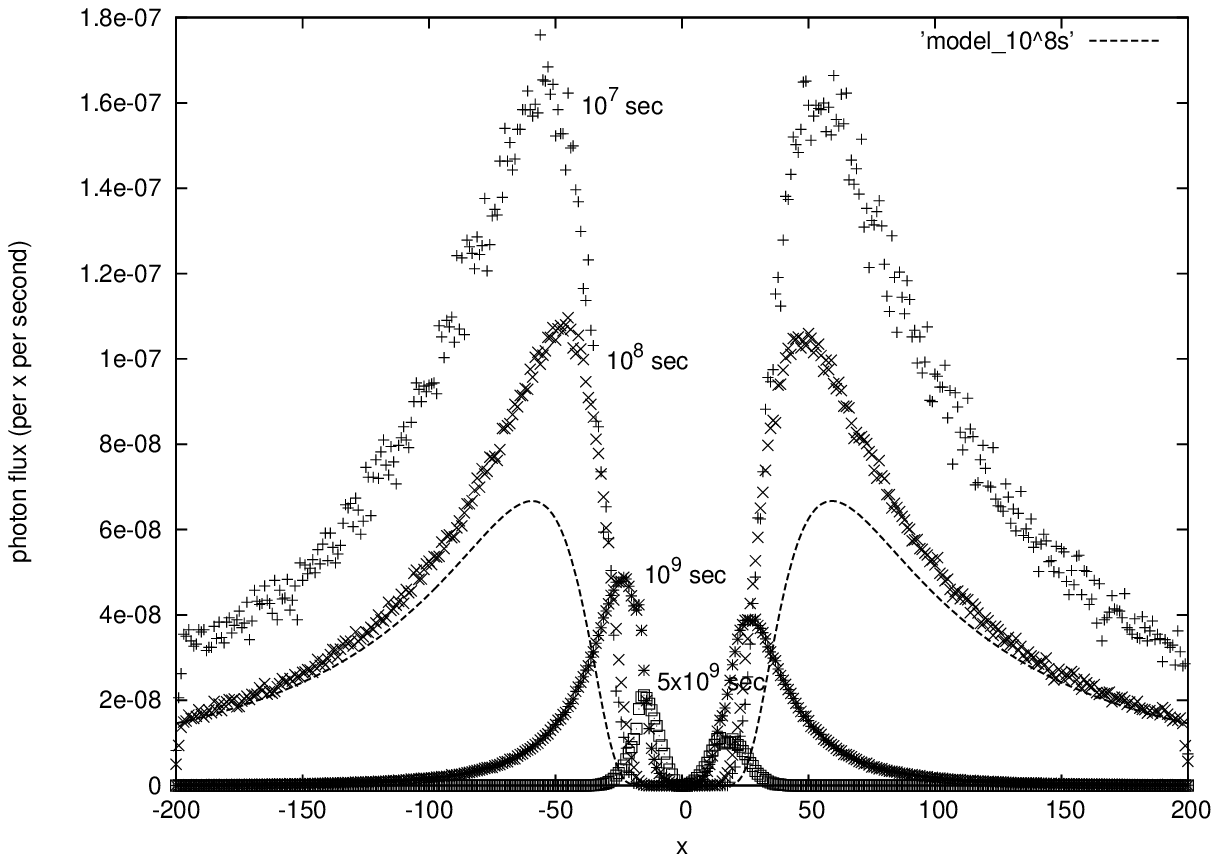}
\caption{ Evolution of emission profile for synthesized spectrum of a GRB afterglow. Synthesized MC simulation results are shown as symboled points for ages of $10^7, 10^8, 10^9 , 5\times 10^9$ sec, respectively from top to bottom for the Sphere model of HI distribution. The dashed line is our analytic model prediction Eq.(11) at age $10^8$ sec. }
\label{fig13}
\end{center}
\end{figure}

Since the radiative transfer equation is linear and the feedbacks of the resonant scattering on the parameters of
 neutral hydrogen are small, 
the total scattered results can be synthesized simply by adding up contributions from sources at different  moments.
Once we have the MC simulation results for a single blasted photon source,  predictions for any kind of source function can be calculated as source weighted integration  over time from the single blast results.
When observation time is large enough, GRB afterglow itself is like a single flashed source. Thus we expect that the synthesized result is not much different from  a single blasted one if its flux normalization is properly adjusted to the duration of the actual burst.

In Fig. 13, we show the synthesized MC results for light evolution in the Sphere scattering model of GRB afterglow. The luminosities are drawn to scale. We see a decaying and narrowing evolution. The MC simulation results agree very well with the analytic model predictions on the wings till $t=10^8$ sec for $x >150$. Yet the discrepancy becomes larger for emissions  closer to the line center. The underestimate is about a factor of 2 for our analytic model in the flux at the peak position. This justifies our estimate on the formation of Lyman $\alpha$ emissions employing the analytic method which includes photons 
 scattered only once (Figs. 2 and 3). The analytic method gives the overall luminosity predictions
of Lyman $\alpha$ emission accurate to order of magnitude. Since these predictions on luminosity are lower bounds, predictions of observability are valid. However, the analytic method predicts poorly on the profile evolution. It doesn't predict a narrowing of the profile as shown in MC simulations. Only  MC simulations reveal the spectral characteristics of the emissions formed by Lyman $\alpha$ scattering.

\section{Discussions and Conclusion}

  Our MC simulation is a complete treatment on the resonant scattering of photons with HI atoms near Lyman $\alpha$ frequency including recoil and frequency transfer.
Yet it turns out that the scattered emission is dominated by photons which are scattered only once. This is true immediately after the GRB's burst , over a period of time
short compared to the cloud's light crossing time, but long compared to GRB's burst duration. This somehow  justifies a perturbational approach (Eqs(7-10)) on the escaping problem of the continuum  light of GRB's afterglow 
 near Lyman  $\alpha$ resonance center.  The directly transmitted flux is the largest component, much larger than all the photons scattered (by $10^4-10^9$ orders of magnitude). Next to it, scattered-once photons take a high percentage of of all  the scattered photons.
As time passes on, scattered twice photons become more numerous than those scattered-once (Fig. 11a). Shortly after that, photons scattered multiple times take 
the domination.
But analogy to a perturbational approach stops here, this is simply a time sequence effect because photons with more scatterings come out later. we don't expect that the  same decreasing ratio between two neighboring terms applies to the higher order terms when the number of scatterings is larger than 2.  At later times when more scatterings happen, the blending of their contribution increases.  Photons experiencing different number of scatterings may take similar percentage in the contribution to the scattered light.

(i) scattered-once photons

It is worthwhile to note that the problem of ``continuum scattering at Lyman $\alpha$ resonance" which
we studied in this paper is different from the problem of ``Lyman $\alpha$ resonant scattering" for photons which are released exactly at Lyman $\alpha$ frequency.
As found earlier by Osterbrock (1962) and Adams (1972), most of the collisions in the second
problem is the ``core scattering". In these problems, the photon escaping is determinated by rare events, either by a single flight or a single excursion depending on the value of optical depth. The escape 
typically takes $(1.5\sqrt{\pi}-1)\tau t_f $  time to escape (Adams 1972).
In the case of GRB's afterglow,
 these ``core scattered" photons do exist, but they originate from  a narrow section of continuum spectrum ($x\sim 1$), and they take a longer time to escape. So their intensity will be  much harder to  to detect. 
In our problem, photons are  ``wing scattered". Photon source is the whole section of continuum ($x\sim100$) up to the frequency where optical depth is of order one. The  escape time scale is the light crossing time scale of the cloud.  Time delay of a scattered and escaped photon is mainly determined by the geometry of the photon's trajectory (Fig.1), rather than the diffusion in frequency space.  The size of the cloud affects more on the predicted luminosity than the hydrogen content in it does. Certainly the cloud has to be rich in HI so that its wing scattering can become effective.

 For the same reason, the escaped photons which are scattered only a few times, or less than one hundred times constitute mostly the escaped luminosity at times when the scattered-once photons finish their role(e.g. Figs.7 and 11a).   The profile difference of MC simulation from analytic model (e.g. Figs. 4b, 6 and 10), which was affected by photons experiencing multiple scatterings,  is actually caused by photons experiencing only up to a few  hundred times.  Photons with millions of times of collision have very little effect on the flux in Fig. 7 and Fig.11a.

(ii) intensity of Lyman $\alpha$ emission 

Separate from the analytic model and the MC simulations, we can estimate the flux of  scattered photons in a simple way: $\frac{\Delta N}{\Delta t}$. The total amount of photons $\Delta N$ released 
are the flux of continuum light times the  frequency width which corresponds to $\tau \sim 1$ 
, say $x=100$, multiplied by
the effective life time of the source, say 10 seconds. These photons are then spreaded over the crossing time of the cloud to escape, say $10^9$ sec ($\sim 10 $pc).
This leaves us with an intensity of scattered light of about $10^{-6}$  of that of the maximum continuum of the observed optical afterglow.

(iii) effects of $\tau$

The column density of neutral hydrogen will affect the total optical depth of the cloud, thus determines the width of the damping troughs by Eq.(6).  Since the emission photons originate from the continuum, a larger optical depth will cut a larger chunk of the continuum to become the source of the scattered photons. The amount of photons (or the equivalent width of the emission) is then approximately proportional to $x_{peak}$, or $\propto \sqrt{\tau}$. But the height of the peaks is little affected because the width of the emission peak scales with $x_{peak}$.

(iv) effects of motion

The gas around GRB may be in motion, as a result of light pressure, gravity, or by turbulence. A typical speed of such motions is that of the sound speed. We considered a simplest model of motion in which all the neutral gas move at a same speed of $V_D$ inward along radius. The bulk speed affects resonant scattering in three ways. First, the neutral hydrogen restframe frequency changed thus the effective optical depth for a given incoming photon has changed. Second, an outcoming photon gets a frequency decrement when they exit in the radial direction by Doppler effect. Third, the differential speed field of the medium causes  an asymmetry of the twin peaks. This effect by velocity gradient is somehow similar to that caused by recoil. From Eq.(3) for each collision, the ratio of frequency change due to velocity gradient to that by recoil is of order $\frac{\delta v/V_D}{b\tau}$, where $\delta v$ is the velocity difference across the cloud. The velocity gradient may become more important than recoil to affect the profile's symmetry if $\delta v >V_D b \tau$.

Fig. 12 shows the effects of motion on the emission profile in the Shell model for a single flashing photon source. In an inward bulk motion, the emission center is redshifted by an amount $ x \sim \frac{u}{V_D}= 1$. The effects on the profile is to make blue peak higher and wider for an inward falling.
 Similar effects will happen for the red peak for an outward flow.
Our findings are in agreement with earlier researches. Urbaniak \& Wolfe (1981) first considered the effects of relative velocity between the Ly$\alpha$ source and the gas. They found that the blue peak is suppressed if the two slabs are recessing from each other; Dijkstra et al . (2006) find similar effects that collapsing gas enhances the blue peak in the escaped flux. 

(v) effects of cloud size and HI distribution

The cloud size, or more exactly the light crossing time, determines the characteristic time scale. Our results for the 5 pc cloud can be scalable
 into other sizes because the radiative transfer equation is linear with time.  For a given cloud size, the result will be different for different distribution of HI. The result of a polytrope distribution is somehow equivalent to changing a uniform cloud into a different size. (Fig.2)   

So far, the existence of clumps of neutral material ( n $\approx 10^3-10^4$ cm$^{-3}$) 
associated with HII regions has been confirmed by CO molecular lines (Williams et al. 1995), and by Si II fine structure lines (Vreeswijk et al. 2004). Clumps with density up to  $10^3$ cm$^{-3}$ are found in environments like Rosette Nebula (Tsvilev et al. 2002). However, these do not exclude clumps with higher density, which could be detected by future molecular line observations. So, the luminosity of the scattered component can be anywhere from  $10^{-4}$  to $10^{-9}$ of the maximum of the transmitted optical flash.

(vi) best time to observe

 If not obscured by host galaxy's DLA or intergalactic neutral hydrogen,
 the time window of observability is 
 from several hours to several years when photons scattered once or twice dominates. The exact time depends highly on the neutral density of the GRB's immediate environment (Figs 2\&3; Figs 7\&11). The intensity prediction can be boosted by several orders of magnitude if the GRB resides in high density filaments or cloudlets.

(vii) distinguishing from other Ly$\alpha$ sources
 
Lyman $\alpha$ emission feature formed by our mechanism has characteristics on its
time variance. The frequency offset of peaks shrinks. So does the width of the peak.
The amplitude may vary.  In the Sphere model, everything changes monotonically. In the Shell model,
the intensity of the peaks may have a second brightening when the photons from the far side of GRB arrive.
The typical time scale for spectral variance is that of the light crossing time of a hydrogen clump, typically one parsec or smaller,
unlike QSO-DLA or galaxy GRB-DLA  which involves kilo-parsec length scale. Therefore,
 the scattered GRB emission can be separated from
 those of the host galaxy  by this time variability, as well as by their spatial compactness.
If observed, the resonant peaks' time dependent behavior acts like a scanning directly on the distribution of neutral hydrogen
in GRB's  immediate neighborhood because photons which arrive at different times correspond to scatterings at different off-sight line distances (Fig.1b).

\acknowledgments

This work is supported
by the Ministry of Science and Technology of China, under
Grant No. 2009CB824900.

\appendix

\section{Method of generating speed $v_z$ of HI atoms}

By introducing an auxiliary parameter $u_0$ and calculating an additional variable $\theta_0$, 
ZM02 algorithm reduces the waste of the exponential suppress at large $v_z$ on the wing scatterings for the rejection method, but introduces computational overhead for core scattering when x is small. Besides, for extremely large x, their algorithm is still inadequate to overcome the rejection waste intrinsic in the method. Thus we improve the algorithm by treating the speed generation differently at different x.

(1) For small $x$ (we adopt as $x<0.6$), since the peaks of $e^{-v_z^2}$ and $\frac{1}{(x-v_z)^2+a^2}$ are very close to each other,the percentage 
waste of rejection is very small. Methods of plain rejection (not employing ZM02's algorithm) is faster because it doesn't have the overheads.

(2) For medium to large $x$ (we adopt as $0.6 \leq x \leq 17$), we basically follow ZM02's algorithm except that we tactfully use $\tilde{p}=1-p$ instead of $p$ for  proper representation of a small number on computer and we set $\theta_0$ as a constant. We treat $u_0$ as a variable which needs to be calculated at each step.

(3) For very large $x$ (we adopt as $x > 17$),  our treatment for $v_z>u_0$ is similar to ZM02. Yet for $v_z\leq u_0$, we switch the roles of the two functions,
using the distribution function $e^{-v_z^2}$ as the transformation method 
to generate $v_z$, and then use $\frac{1}{(x-v_z)^2+a^2}$ as the comparison function to reject. This is more effective because for large $x$, the Lorentz function is a slow varing function while the Gaussian function decays fast.

%\end{thebibliography}

\end{document}